\documentclass[11pt]{article}

\RequirePackage[OT1]{fontenc}
\RequirePackage{amsthm,amsmath,amssymb}
\RequirePackage[numbers]{natbib}
\RequirePackage[colorlinks,citecolor=blue,urlcolor=blue]{hyperref}
\RequirePackage{bm,graphicx}

\usepackage[dvipsnames]{xcolor}

\usepackage{multicol}

\def\FIGS{}
\title{High-resolution Bayesian mapping of landslide hazard with unobserved trigger event}

\author{Thomas Opitz$^1$, Haakon Bakka$^2$, Rapha\"el Huser$^3$, Luigi Lombardo$^4$ \vspace{.2cm}\\
	$^1$ Biostatistics and Spatial Processes, INRAE, Avignon, France\\
	$^2$ University of Oslo, Oslo, Norway \\
	$^3$ Computer, Electrical and Mathematical Sciences and Engineering (CEMSE) Division,\\ King Abdullah University of Science and Technology (KAUST), Thuwal 23955-6900, Saudi Arabia\\
	$^4$  Faculty of Geo-Information Science and Earth Observation, University of Twente, Netherlands
}

\begin{document}
	\maketitle

\begin{center}
	\begin{minipage}{0.9\linewidth}
		{\textbf{Abstract. }}
		
Statistical models for landslide hazard enable mapping of risk factors and landslide occurrence intensity by using  geomorphological covariates available at high spatial resolution. However, the spatial distribution of the triggering event (\emph{e.g.}, precipitation or earthquakes) is often not directly observed. In this paper, we develop Bayesian spatial hierarchical models for point patterns of landslide occurrences using different types of log-Gaussian Cox processes. Starting from a competitive baseline model that captures the unobserved precipitation trigger through a spatial random effect at slope unit resolution, we explore novel complex model structures that take clusters of events arising at small spatial scales into account, as well as nonlinear or spatially-varying covariate effects. For a $2009$ event of around $4000$ precipitation-triggered landslides in Sicily, Italy, we show how to fit our proposed models efficiently using the integrated nested Laplace approximation (INLA), and rigorously compare the performance of our models both from a statistical and applied perspective. In this context, we argue that model comparison should not be based on a single criterion, and that different models of various complexity may provide insights into complementary aspects of the same applied problem. 
In our application, our models are found to have mostly the same spatial predictive performance, implying that key to successful prediction is the inclusion of a slope-unit resolved random effect capturing the precipitation trigger.  Interestingly, a parsimonious formulation of space-varying slope effects reflects a physical interpretation of the precipitation trigger: in subareas with weak trigger, the slope steepness is shown to be mostly irrelevant. \vspace{.2cm}

	\end{minipage}
\end{center}

\section{Introduction}

%\ \\Main goals:
%\begin{enumerate}
%	\item estimate complex model for sparse discrete data in high dimensions: log-Gaussian Cox processes (INLA our tool of choice)
%		\item special focus 1: overdispersion (Poisson process : nonstationaritities and random effects), with $K$-function as analysis tool
%		\item special focus 2: unobserved trigger and its interaction with effects of observed covariates
%	\item model selection and validation for complex log-Gaussian Cox processes (cross-validation and prediction scores)
%
%\end{enumerate}

%1) Some background on predictive landslide modeling
Most of the current approaches to mapping landslide hazard exploit auxiliary information from geomorphological covariates and focus on one of its components---known as the landslide susceptibility---through the modeling of presence-absence information \citep{atkinson1998generalised,ayalew2005application,Camilo2017,Goetz.al.2015,rossi2010optimal}; see \citet{Reichenbach.al.2018} for a recent review. Such approaches  are predominantly based on machine learning techniques using binary classification over fine pixel grids, where subsampling of zero observations is often inevitable to cope with high spatial resolution and highly imbalanced designs. Recently, \citet{Lombardo.etal:2018,lombardo2019numerical} and \citet{lombardo2019geostatistical,lombardo2019space} introduced the ``intensity'' concept for spatial landslide prediction by focusing on event counts and not only presence-absence data. Specifically, they proposed a  novel probabilistic approach based on a Bayesian hierarchical models, where landslides are viewed as spatial or spatio-temporal point processes of log-Gaussian Cox type.  Using the integrated nested Laplace approximation \citep{Illian.al.2012,Rue.Martino.Chopin.2009}, they developed accurate statistical inference with high grid resolution and with sophisticated latent structures for capturing intensity variations not explained by observed covariates. In this paper, we use one of their models as a highly competitive baseline, and explore various more complex model extensions described below. While observed covariates are available at high pixel resolution, spatial random effects can be resolved at multiple scales. Here, we use \emph{slope units} \citep[SUs,][]{amato2019,carrara1995gis}, which allow fast computations by being at lower spatial resolution than pixels, while delimiting physically-motivated zones that are relevant to the landsliding process, which is known to show a relatively homogeneous response 
%\BakkaComment{What is response here? Maybe write out in another sentence?} 
to slope instabilities within each SU. SUs are  commonly used in landslide science (and more generally geomorphology), because of the empirical evidence that landslides occur on slopes \citep{guzzetti1994towards}. %\BakkaComment{Drop all words like indeed: too informal for AOAS.}

We study non-trivial model extensions with respect to two important aspects: first, we include spatially unstructured effects at pixel or SU scale to capture residual spatial clustering at small scales; second, we construct complex models with nonlinear or space-varying covariate effects, in order to improve the baseline model's predictive performance and allow for new insights and interpretations from an applied perspective. Space-varying regression has been established as a useful concept when the response to a covariate is not stationary in space \citep{Gamerman.al.2003,Gelfand.al.2003}. In our context, we consider it as a natural---yet difficult to implement---solution to account for the spatially-varying influence of the landslide trigger (such as a heavy precipitation event), which is usually not observed at good spatial resolution in the study area, or not observed at all. In this paper, we conduct a thorough and rigorous structural analysis based on models that explore how the unobserved trigger interacts with observed covariate effects, here highlighted with the example of the slope steepness. Throughout our statistical analysis, our aim is to integrate physical understanding of the landsliding process into the model structure, e.g., by assuming that the slope steepness should become irrelevant as a predictor in places where there is only a weak---or no---trigger influence. Such improvements have never been considered in the landslide modeling literature, despite being of high practical relevance. More generally, we also discuss diagnostics to comprehensively compare the goodness-of-fit and spatial predictive performance of a range of models of varying complexity, in order to study improvements with respect to the baseline. We argue that decisions should not be based on a single criterion, but rather on a combined assessment of several criteria to more objectively appreciate the relative strengths and weaknesses of models. Different models may in fact give complementary insights into the statistical or physical behavior of landslide activations.
%\TC{(Raphael: Should we mention here that different models give complementary insights into the statistical or physical behavior of landslide activations?)}

In the remainder of the paper, Section~\ref{sec:data} describes the landslide data and geomorphological covariates that we use in our analysis.  Then, Section~\ref{sec:modeling} provides the necessary background theory on log-Gaussian Cox processes, and presents our new models, while Section~\ref{sec:inference} discusses implementation and model comparison using the integrated nested Laplace approximation. In Section~\ref{sec:results},  we compare and interpret the fitted models, which we exploit to map the landslide intensity and slope-related risk factors.  We conclude with some discussion in  Section~\ref{sec:discussion}.

\section{Landslides data and predictor variables}
\label{sec:data}
\subsection{Precipitation-triggered landslide occurrences}
\begin{figure}[t!]
\includegraphics[width=.475\linewidth]{\FIGS 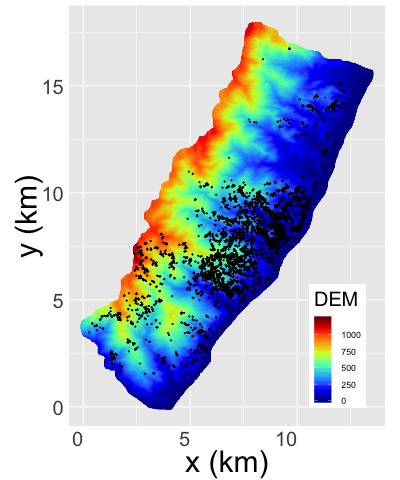} \quad 
\includegraphics[width=.475\linewidth]{\FIGS 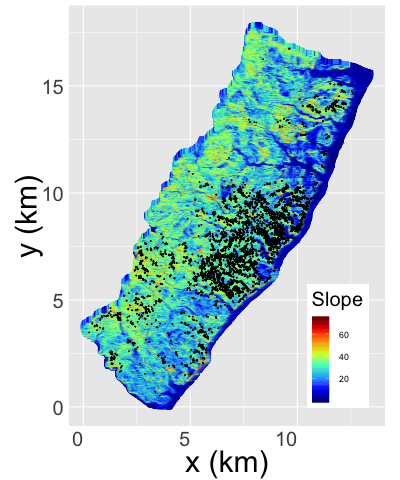}
\caption{Study area. The maps show the digital elevation model (DEM, left), the Slope Steepness (right), as well as the landslide inventory (with black dots representing the LIPs).
%with size proportional to the number of landslides observed in the corresponding pixel). 
}
\label{fig:data}
\end{figure}

On October 1, 2009, a major rainfall discharge occurred in an area of around $100$~km$^2$ on the island of Sicily (Southern Italy), with approximately $250$~mm of rain measured  at nearby weather stations. This weather event followed two relatively smaller precipitation events  just one and two weeks before, with about $190$~mm and $75$~mm of rain, respectively \citep{cama2015predicting,lombardo2016a}. Within just a few hours, this extreme precipitation event triggered  several thousands of rapid shallow landslides and led to the death of $37$ people, and economic infrastructure damage of around half a billion Euro.  Using remote sensing images before and after the event, the identification of $4874$ separate debris-flow landslide \citep{hungr2014varnes} was made possible. 
%\citet{Lombardo.etal:2018} developed a first high-resolution model of  the whole area using a   log-Gaussian Cox process model. 
So-called landslide identification points \citep[LIPs,][]{lombardo2014test} were then extracted from remotely sensed images (at $0.25$m resolution) for each mass movement. Precisely, the triggering location was set to the point of highest altitude in the area affected by the movement. The left panel of Figure~\ref{fig:data} shows a digital elevation model of the study area and the LIP inventory at $15$m pixel resolution. 
%, where the surface area of points is proportional to the number of landslides observed in a 15m $\times$ 15m squared lattice. 
Most LIPs were recorded in distinct pixels, but with a few exceptions: 353 pixels contained 2 landslides, 44 pixels contained 3 landslides, and 2 pixels contained 4 landslides.    % and the division into $12$ so-called macro-catchments.
%\TC{Have to present a ``new" figure here. We could show a big figure of slope, too, since we focus on slope for the space-varying regressions.}

\subsection{Geomorphological covariate information}

We use covariate information that has been aggregated to a $15$m~$\times$~$15$m grid based on a  digital elevation model (DEM) with $2$m resolution. This  grid resolution has been shown to be sufficiently fine to avoid degrading the predictive performance of models in this study area \citep{arnone2016effect,cama2016exploring,lombardo2016b}, while allowing for reasonably fast inference. \citet{Lombardo.etal:2018} provide a more detailed description of the calculation and meaning of covariates.  Aggregation from $2$m to $15$m was done by averaging values for continuous covariates, or by selecting the prevailing category for categorical covariates. Covariates are as follows, where we use upper-case notation throughout to refer to the names of these covariates: Elevation (or Digital Elevation Model, abbreviated DEM); Aspect, i.e., the angle in $[0,2\pi)$ describing the exposition of the area with respect to the North \citep{zevenbergen1987quantitative}; Slope Steepness \citep{zevenbergen1987quantitative}; Planar Curvature \citep{heerdegen1982quantifying}, which is measured perpendicular to the steepest slope angle and characterizes the convergence and divergence of flow across the surface; Profile Curvature \citep{heerdegen1982quantifying}, which indicates the direction of maximum slope; Topographic Wetness Index (TWI) \citep{beven1979physically}, which  quantifies topographic properties related to hydrological processes using slope and upstream contributing area as input; Stream Power Index (SPI) \citep{moore1991digital}, which takes similar input as TWI and measures more specifically the erosive power of flowing water; Landform (with $10$ categories; see \citet{wilson2000digital}); the distance of each pixel to the closest tectonic fault line (Dist2Fault); Normalized Difference Vegetation Index (NDVI) \citep{rouse1974monitoring}, which measures the ``greenness'' of a landscape and serves as a proxy for vegetation; Lithology, i.e., soil type with $22$ categories, where rare soil types with less than $500$ occurrence pixels have been summarized in a single class ``other'';  Land Use (with $13$ categories). The choice of a $15$m~$\times$~$15$m grid yields a representation of the study area through $449,038$ pixels. %; the dimension reduction with respect to the original $2$m resolution did not affect model performances in several applications in the same study area \citep{arnone2016effect,cama2016exploring,lombardo2016b}. 
When using continuous covariates for modeling purposes, we scale them to have empirical mean $0$ and empirical variance $1$. Additional information about the covariates can be found in Section~1 of the Supplementary Material.   %\TC{(Raphael: Say that continuous covariates have been standardized to have mean zero and variance one? (seems important for the prior specification below, and for interpretation...))}

 Figure~\ref{fig:data} shows the spatial distribution of the Elevation (left panel) and the Slope Steepness (right panel) on their original scale. 
 %\TC{(Put maps of other covariates in Supplementary Material? Raphael: Perhaps that would be good, I think. But perhaps more importantly it would be good to have a very brief explanation of all covariates, as planar curvature or TWI might be totally unknown to statisticians...)} 
 In this work, we stress the importance of accurately capturing the influence of the Slope Steepness, which has a major effect on landslide activations.
 %, and throughout we use the upper-case notation ``Slope" to refer to the values of this covariate. 
Landslides are very unlikely to be triggered in flat areas; they are much more likely on steeper slopes; and they become unlikely again on very steep slopes, since movable material has already gone through erosion and during previous mass movements. % and due to continual weather influence. 
Below, we seek to construct models that can capture this non-monotonic and highly nonlinear influence of the Slope Steepness. 

\begin{figure}[t!]
\includegraphics[width=.95\linewidth]{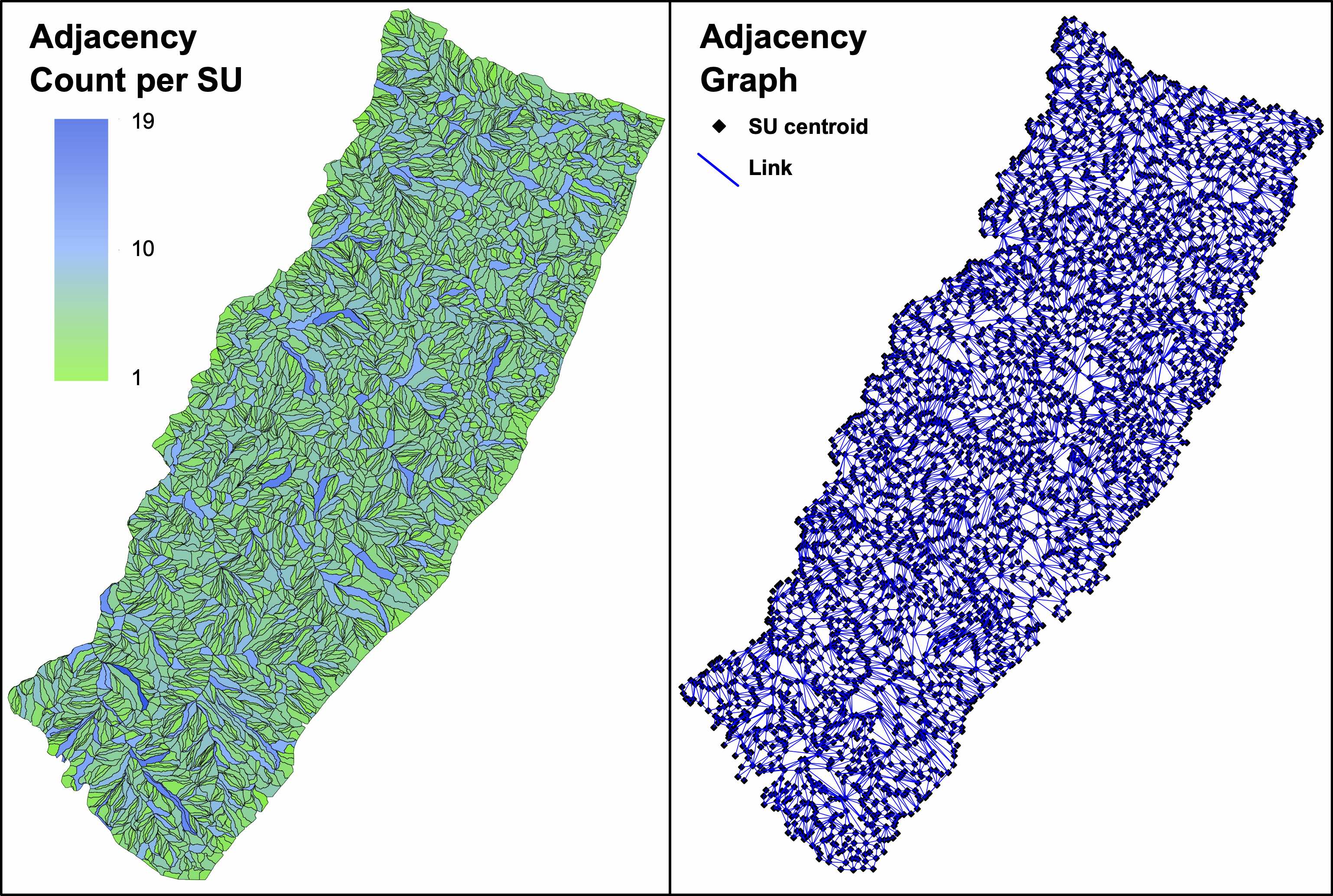} 
\caption{Illustration of slope units (SUs) and of their adjacency structure. Left:  Number of adjacent SUs to each SU indicated by color. Right: Adjacency graph. }
\label{fig:su}
\end{figure}

In our Bayesian hierarchical models described in Section~\ref{sec:modeling}, we additionally exploit modeling units at an intermediate resolution (between $15$m-pixels and the full study area) known as \emph{slope units} (SUs) \citep{alvioli2016automatic,Lombardo.etal:2018}. Precisely, the SU partition defines a phyically-motivated and moderately-sized spatial discretization, here used for capturing latent random effects such as the influence of the spatially-varying precipitation event. These SUs can be viewed as relatively homogeneous mapping units with respect to geomorphological and geophysical features that are relevant to landslide activations. In our study area, we have $3484$ SUs, which are displayed on the left panel of Figure~\ref{fig:su}.
%For an intermediate resolution of modeling units (between $15$m-pixels and the full study area), we utilize a subdivision into $3484$ \emph{slope units} (SUs) \citep{alvioli2016automatic,Lombardo.etal:2018}, which define a phyically-motivated and moderately-sized spatial discretization for capturing latent random effects such as the influence of the spatially-varying precipitation event. These SUs can be viewed as relatively homogeneous mapping units with respect to geomorphological and geophysical features that are relevant to landslide activations. The configuration of SUs for our dataset is shown on the left panel of Figure~\ref{fig:su}. 
Some landslides triggered within the same SU may be due to a joint triggering mechanism, which may potentially lead  to  some residual stochastic dependence in the landslide occurrence process (conditional to the geomorphological structure and the precipitation trigger), but such spatial dependence is likely to be very weak when considering events arising in separate SUs. In other words, landslide data in different SUs can be safely assumed to be conditionally independent given fixed and random effects. In the next section, we describe our proposed log-Gaussian Cox process models.
%Our modeling assumption is that landslide events occur independently conditional to  the geomorphological covariates and  the precipitation event,  which will enter our models through fixed and latent random effects. 

%\section{Bayesian hierarchical modeling with log-Gaussian Cox processes}
\section{Bayesian hierarchical modeling of landslide point patterns}
\label{sec:modeling}

\subsection{Log-Gaussian Cox processes}

Spatial point processes are stochastic models for the occurrence of events in space, when the event positions are random but obey certain density patterns and small-scale clustering or inhibition behavior. %We restrict our attention to simple point processes without accumulation points, such  that observing two points at the same site $s$ or at two sites at infinitesimally small distance is an impossible event, which is an appropriate assumption for modeling landslides data. 
We refer to the bounded study area as $\mathcal S$, shown in Figure~\ref{fig:data}. Point processes characterize the joint probability distribution of the number of points $N(A_i)$ in sub-areas  $A_1,A_2\ldots$, i.e., the marginal distribution and dependence structure of the random count variables $N(A_i)$. An important characteristic of point processes is the intensity function $\lambda(\bm s)$, which determines the expected number of points over any area $A\subset \mathcal S$, i.e., $\mathbb{E}\{N(A)\}=\int_A \lambda(\bm s)\, \mathrm{d}\bm s$. For Poisson processes, $N(A)$ has the Poisson distribution for any set $A$, and the occurrence of points is independent given the (deterministic) intensity function $\lambda$.

Log-Gaussian Cox processes  \citep[LGCPs,][]{Moller.al.1998} are Poisson processes with a stochastic intensity function $\Lambda(\bm s)$ given by a log-Gaussian process. Their doubly-stochastic structure allows capturing spatial clustering of points due to unobserved or unavailable predictor variables. 
LGCPs are convenient for Bayesian hierarchical modeling, where the latent Gaussian process may encompass both fixed effects of observed covariates $z_j(\bm s)$, and random effects.

Throughout, we use the notation $x(\bm s)$ and $\bm x$ for random effects by adding context-specific subscripts and superscripts. Precisely, we follow the convention that $x(\bm s)$ corresponds to the value of a random effect evaluated at the location $\bm s\in\mathcal S$, and we write $\bm x$ for the vector of the finite number of latent variables (following a multivariate Gaussian distribution) that are used to represent this random effect. For example, a random effect $\bm x_{\mathrm{SU}}$ described at SU resolution corresponds to a multivariate Gaussian random vector with $3484$ components, one for each SU, while $x_{\mathrm{SU}}(\bm s)$ is the value that corresponds to the SU containing the location $\bm s\in \mathcal S$.

%In general, the log-intensity  of our models  possesses the following additive structure
The log-intensities of our models are structured as
\begin{equation*}\label{eq:gauss}
\log \Lambda(\bm s) = \beta_0+\sum_{j=1}^J \beta_j z_j(\bm s) + \sum_{k=1}^K x_k(\bm s), \quad \bm s\in \mathcal S.
\end{equation*}
The random effects $x_k(\bm s)$ may directly depend on location $\bm s$, or only indirectly through a covariate $z_j(\bm s)$ observed at $\bm s$, for instance  if $x_k(\bm s)$ is used to capture the potentially nonlinear influence of a covariate. 
%Here,  $W_k$ encodes random effects  defined with respect to an index $\bm z_k$ e.g., pixel, slope unit, or classes of values of a covariate), and is required to be be multivariate Gaussian. 
The probability density function of an observed finite point pattern $\mathcal X = (\bm X_1,\ldots, \bm X_N)^T$, composed of a random but finite number  $N\geq 0$ of points $\bm X_i\in\mathcal S$ in the observation window $\mathcal S$, corresponds to the expectation 
\begin{equation*}\label{eq:lgcpdens}
f_{\mathrm{LGCP}}( \mathcal{X})=\mathbb{E}_{\Lambda}\left[ \exp\left(-\int_{\mathcal S} \Lambda(\bm s)\,\mathrm{d}\bm s\right) \prod_{i=1}^N \Lambda(\bm X_i)\right],
\end{equation*}
using the convention that $\prod_{i=1}^N \Lambda(\bm X_i)=1$ if $N=0$. 
Closed-form expressions of this expectation %of the integral with respect to the latent log-Gaussian process  $\Lambda(s)$ 
are not available in general, but Bayesian inference techniques, such as  implementations based on the integrated nested Laplace approximation \citep{Rue.Martino.Chopin.2009} as used here, have been developed to approximate it numerically. In our Bayesian framework, the Gaussian processes used to construct the log-Gaussian intensity function in LGCPs can also be viewed as prior distributions for deterministic components of the intensity function of a Poisson process. %to estimate the intensity function $\Lambda(s)$   and hyperparameters related to its structure without  performing direct numerical integration in \eqref{eq:lgcpdens}. Therefore, Gaussian prior distributions are specified for all components of $\log \Lambda(s)$ to be estimated, including the fixed effects $\beta_j$, $j=0,\ldots,J$.  The Gaussian process in \eqref{eq:gauss} then defines the prior distribution for the log-intensity of a Poisson process. 

\subsection{Models}

We now present the baseline model and our proposed model extensions. Throughout, we make systematic use of the concept of penalized complexity (PC) priors \citep{Simpson.al.2017} to define priors for hyperparameters (e.g., for standard deviations of random effects). For each latent component of the model, a simple baseline model is defined, typically corresponding to its absence (e.g., a standard deviation of $0$) such that all the latent  variables pertaining to the component are set to  $0$. PC priors are then defined through a constant-rate penalty for  the distance to the baseline, expressed in terms of a transformation of the Kullback-Leibler divergence. In particular, the PC prior of precision parameters corresponds to an exponential prior on the corresponding standard deviation \citep{Simpson.al.2017}.     
%\TC{discuss penalized complexity priors}

\subsubsection{M0 (Baseline): Fixed effects and spatial random effect}
%\subsection{Integrating geomorphological data}
%splines+lincomb vs. iid?
%number of levels: follow common practice in the literature for landform (10)
%slope units are homogeneous, and there may be dependence / overdispersion within a slope unit
%after geological and land use covariates, only triggering weather event remains t
%o explain spatial dependence between slope units

% Model 21

Our baseline model, called M0, has structure similar to the best model found by \cite{Lombardo.etal:2018},  although here with relevant improvements through the choice of  prior distributions penalizing model complexity. In the model extensions presented subsequently,  we ensure easy comparison and consistency with M0 by keeping the same prior distributions for components that are in common. 
%Overall, we fix certain hyperparameters to appropriately chosen values to reduce the computational cost of INLA and to ensure  stability of the estimation algorithm; our choice of fixed values is based on preliminary experiments for exploring the stability of estimates and convergence of the estimation algorithm. Owing to the relatively small number of latent variables in  the  effects with fixed hyperparameters and the relatively large sample size, we expect posterior uncertainty  to be low for such effects, and the risk of oversmoothing due to too informative priors is not a critical issue. 
%Using notation $\mathrm{SU}(s)$ to denote the index of the SU containing the location $s$, 
Before giving details on components of the log-intensity of M0, we provide its full formula:
\begin{equation}
\log \Lambda_0(\bm s)=\sum_{j=1}^{m_{\mathrm{cont}}} \beta_j^{\mathrm{cont}}  z^{\mathrm{cont}}_j(\bm s)  +  \sum_{j=1}^{3}\sum_{\ell=1}^{\ell_{j}} \beta_{j,\ell}^{\mathrm{cat}}  z_{j,\ell}^{\mathrm{cat}}(\bm s) +  x_{\mathrm{Aspect}}^{\mathrm{CRW1}}(\bm s) +  x_{\mathrm{SU}}^{\mathrm{CAR}}(\bm s)\tag{M0}.
\end{equation}

The intercept and the continuous covariates, $z^{\mathrm{cont}}_j(\bm s)$, including the Slope Steepness but with the exception of the circular Aspect variable, appear as fixed effects with coefficients $\beta^{\mathrm{cont}}_j$, $j=1,\ldots,m_{\mathrm{cont}}$. To guide the estimation algorithm for faster convergence and to stabilize the fitted model, we fix moderately informative Gaussian prior distributions with precision $1$ and mean $0$  for fixed effects, except for the intercept where the mean is $-2$.  

PC priors are used for the precision of the independent and identically distributed (i.i.d.) effects of the three categorical, non-ordinal covariates (Lithology, Landform, Land Use), $z_{j,\ell}^{\mathrm{cat}}(\bm s)$, which possess a substantial number of categories ($\ell_1=22$, $\ell_2=10$ and $\ell_3=13$, respectively). Priors for their coefficients $\beta_{j,\ell}$,  $\ell=1,\ldots,\ell_j$, $j=1,2,3$, are centered at $0$, and we impose a sum-to-zero constraint on the coefficients of each of the three factors to ensure identifiability. The priors for the three precision parameters, denoted by $\tau_{\mathrm{LITH}}$, $\tau_{\mathrm{LF}}$ and $\tau_{\mathrm{LU}}$, respectively, are relatively informative, and are determined by the a-priori specification of $\mathrm{Pr}(\sqrt{1/\tau}>1) = 0.01)$. Therefore, we let the data drive the posterior distribution away from $0$ if a clear signal is present in the data. This allows us to shrink the model towards a parsimonious formulation and to avoid unstable inference.

The Aspect covariate, $x_{\mathrm{Aspect}}^{\mathrm{CRW1}}(\bm s)$, reports an angle within the interval $[0,2\pi)$, which we here discretize into $16$ equidistant bins, each spanning $2\pi/16=22.5^\circ$ for a near-continuous treatment. The prior model of this random effect is a cyclic first-order random walk (CRW1)  over the bins, with a sum-to-zero constraint for identifiability. Writing $\bm x_{\mathrm{Aspect}}^{\mathrm{CRW1}}=(x_{\mathrm{A},0},\ldots,x_{\mathrm{A},15})^T$, we characterize its multivariate Gaussian prior through the following conditional specification (where $\tilde{i} = i\ \mathrm{modulo}\ 16$):
\begin{equation*}\label{eq:crw1}
%x_{\mathrm{Aspect},0}=x_{\mathrm{Aspect},16},  \quad
        x_{\mathrm{A},i}\mid \{x_{\mathrm{A},\widetilde{i-1}},x_{\mathrm{A},\widetilde{i+1}}\} \sim \mathcal{N}\left(\frac{x_{\mathrm{A},\widetilde{i-1} }+x_{\mathrm{A},\widetilde{i+1} }}{2},\frac{1}{\tau_{\mathrm{A}}\tau_{A,0}}\right), \ i=0,\ldots, 15,
\end{equation*}
where the constraint $\sum_{i=0}^{15} x_{\mathrm{A},i}=0$ is imposed to ensure identifiability, and
where $\tau_{A,0}>0$ is a fixed scaling constant such that $1/\tau_{\mathrm{A}}$ corresponds to the marginal variance of the variables $x_{A,i}$, $i=0,\ldots,15$. The CRW1-structure makes sure that the estimated piecewise constant curve is ``smooth'' by borrowing strength between neighboring classes. We set an informative PC prior distribution for $\bm x_{\mathrm{Aspect}}^{\mathrm{CRW1}}$ by specifying that, a priori, $\mathrm{Pr}(\sqrt{1/\tau_{\mathrm{A}}}>1) = 0.01$. 

For the latent spatial effect $x_{\mathrm{SU}}^{\mathrm{CAR}}(\bm s)$ structured at the SU level, we implement Besag's classical conditional autoregressive (CAR) model \citep{Besag.1975,Rue.Held.2005}. Writing $\bm x_{\mathrm{SU}}^{\mathrm{CAR}}=(x_{\mathrm{SU},1}^{\mathrm{CAR}},\ldots,x_{\mathrm{SU},3484}^{\mathrm{CAR}})^T$, this model links the value $x_{\mathrm{SU},i}^{\mathrm{CAR}}$
%=f_{\mathrm{SU}}(i)$ 
in SU  $i$ to adjacent SUs, described by the index set $\mathrm{NB}(i)$ of size $n_i\geq 1$, as: %in the following way:
\begin{equation*}\label{eq:spateff}
x_{\mathrm{SU},i}^{\mathrm{CAR}}\mid \{x_{\mathrm{SU},j}^{\mathrm{CAR}}\}_{j\in\mathrm{NB}(i)}\sim \mathcal{N}\left(\frac{1}{n_i}\sum_{j\in\mathrm{NB}(i)} x_{\mathrm{SU},j}^{\mathrm{CAR}}, \frac{1}{n_i\tau_{\mathrm{SU}}\tau_{\mathrm{SU,0}} } \right), \ i = 1, \ldots, 3484,
\end{equation*}
where the constraint $\sum_{i=1}^{3484} x_{\mathrm{SU},i}^{\mathrm{CAR}}=0$ is imposed for identifiability,
%where $\tau>0$ is a precision parameter that governs the spatial variation of the effect. 
and where $\tau_{\mathrm{SU},0}>0$ is a fixed scaling constant such that $1/\tau_{\mathrm{SU}}$ corresponds to the (generalized) marginal variance \citep{Sorbye.Rue.2014} of the variables $x_{\mathrm{SU},i}^{\mathrm{CAR}}$, $i=1,\ldots,3484$, where the generalized marginal variance is the square of the geometric mean of the (nonstationary) standard deviations of the variables. We set a moderately informative PC prior distribution for $\bm x_{\mathrm{SU}}^{\mathrm{CAR}}$ by specifying that, a priori, $\mathrm{Pr}(\sqrt{1/\tau_{\mathrm{SU}}}>5) = 0.01$. 

The number of adjacent SUs, $|\mathrm{NB}(i)|$,  varies moderately in our dataset, with $90\%$ of values between $3$ and $8$ and $60\%$ between $4$ and $6$, while the minimum is $1$ and the maximum is $19$; see Figure~\ref{fig:su}. 
%This prior model may induce higher dependence of a slope unit on the average behavior of its neighbors when their number is higher, i.e., typically when a large slope unit is large or when the  surface relief is locally very rough, leading to a relatively higher number of neighbors. We do not consider such behavior as counterintuitive for our goal of reconstructing the precipitation event. 
In general, this spatial random effect captures SU-resolved effects that cannot be explained by  other model components, in particular by observed covariates. 
%Alternatively, it may also be viewed as a nonparametric covariate effect of spatial position. 
With the landslides data, the spatial effect will absorb the local intensity variation of the precipitation trigger, which is at most weakly correlated with some of the other, geomorphological covariates. 

%Due to the relatively complex prior model with a high number of latent Gaussian variables distributed over various latent effects, we only estimate one hyperparameter, which is the CAR coefficient of the slope unit model characterizing the strength of dependence over space. Its value is of crucial importance for spatial smoothing and interpretation of results, and we choose a moderately informative gamma prior distribution centred at $1$. 

\subsubsection{Model 1: Spatially unstructured  effects}
Overdispersion in count data refers to the situation where the variance is larger than the mean, which stands in contrast to the Poisson distribution whose mean and variance are equal. Conceptually, our LGCP models are defined over continuous space and exclude multiple events at the same location $\bm s\in\mathcal S$, such that (theoretically) the notion of  overdispersion does not apply. However, overdispersion in the counts for spatial units such as pixels or SUs can still arise if our intensity model is misspecified and fails to pick up all sources of spatial variation in the data. Our models assume constant intensity within pixels, and the spatial random effect has coarser resolution at the SU level. Pixel resolution is very high in our case and approximates continuous space,  with only a very small proportion of  pixels counting multiple events, such that we will not explore nonstationary behavior of the point process intensity within pixels.
%study within-pixel overdispersion.
%\BakkaComment{Previous sentence says no pixel iid effect, next sentence says yes to pixel iid effect. Only one can be correct.} 
However, we propose to explore models with spatially unstructured effects at the pixel level or the SU level, which are capable of capturing sharp differences in intensity between neighboring pixels or SUs, respectively.  In our first model extension, we therefore include i.i.d.\ Gaussian variables in the latent linear predictor, either by adding  one variable to each pixel (Model 1a), or by adding one variable to each SU (Model 1b). We estimate the precision parameters, $\tau_{\mathrm{grid}}^{\mathrm{iid}}$ and $\tau_{\mathrm{SU}}^{\mathrm{iid}}$, of these pixel-wise and SU-wise unstructured effects, respectively. Writing ${\rm i}(\bm s)$ and ${\rm SU}(\bm s)$ to denote the pixel ${\rm i}$ and slope unit containing location $\bm s\in\mathcal S$, respectively, these models are given as
%(Model 22 is iid on SUs, model 23 is iid on pixels)
\begin{align*}
\log \Lambda_{1a}(\bm s)&=\log \Lambda_{0}(\bm s)+x_{\mathrm{grid}}^{\mathrm{iid}}\{\mathrm{i}(\bm s)\}, \ \  \bm x_{\mathrm{grid}}^{\mathrm{iid}}\sim \mathcal{N}_{n_{\mathrm{grid}}}(\bm 0,I_{n_{\mathrm{grid}}}/\tau_{\mathrm{grid}}^{\mathrm{iid}}), \tag{M1a}\quad  \\
\log \Lambda_{1b}(\bm s)&=\log \Lambda_{0}(\bm s)+x_{\mathrm{SU}}^{\mathrm{iid}}\{\mathrm{SU}(\bm s)\},  \ \ \bm x_{\mathrm{SU}}^{\mathrm{iid}}\sim \mathcal{N}_{n_{\mathrm{SU}}}(\bm 0,I_{n_{\mathrm{SU}}}/\tau_{\mathrm{SU}}^{\mathrm{iid}}),\tag{M1b}
%f^{iid}_{SU}(s)
\end{align*}
where sum-to-zero constraints are imposed on $\bm x_{\mathrm{SU}}^{\mathrm{iid}}$ and $\bm x_{\mathrm{grid}}^{\mathrm{iid}}$, $\bm 0$ denotes the zero vector, $I_n$ is the $n$-by-$n$ identity matrix, and $n_{\rm grid}=449,038$, $n_{\rm SU}=3484$. For both effects, the  precision parameter $\tau$ ($=\tau_{\mathrm{grid}}^{\mathrm{iid}}$ or $\tau_{\mathrm{SU}}^{\mathrm{iid}}$) is endowed with an % \BakkaComment{weakly instead?} 
informative PC prior determined by $\mathrm{Pr}(\sqrt{1/\tau}>1) = 0.01$. 

\subsubsection{Model  2: Nonlinear effect}
% Model 24

In this second model extension, we replace the linear Slope Steepness effect of the form ``$\beta_{\mathrm{Slope}} \times \mathrm{Slope}(\bm s)$'' (with $\mathrm{Slope}(\bm s)$ a known covariate) by a nonlinear random effect $\bm x_{\mathrm{Slope}}^{\mathrm{RW1}}$, defined through a first-order random walk prior using $10$ equidistant classes to partition Slope Steepness values. Denote the log-intensity of the baseline model without the linear Slope Steepness effect by $\log \Lambda_{0,-\mathrm{Slope}}(\bm s)$. Here, we consider the modified model 
%(Model 24)
\begin{equation}\tag{M2}
\log \Lambda_{2}(\bm s)=\log \Lambda_{0,-\mathrm{Slope}}(\bm s)+ x_{\mathrm{Slope}}^{\mathrm{RW1}}(\bm s),
\end{equation}
which can capture nonlinear, and in particular non-monotonic influence of the Slope Steepness covariate. 
We set the prior distribution of the precision parameter $\tau_{\mathrm{Slope}}^{\mathrm{RW1}}>0$ of $\bm x_{\mathrm{Slope}}^{\mathrm{RW1}}$ by analogy with the Aspect effect in M0, i.e., 
$\mathrm{Pr}(\sqrt{1/\tau_{\mathrm{Slope}}^{\mathrm{RW1}}}>1) = 0.01$. 

\subsubsection{Model 3: Space-varying regression (SVR)}
Another extension of our baseline model is possible by keeping a linear coefficient for Slope but allowing it to vary over space. This allows the model to capture local variations of the strength of the Slope Steepness effect due to the precipitation trigger. We keep the global linear Slope Steepness coefficient and add a spatially-varying correction, defined at the SU level,  in the following model:
%(Model 25)
\begin{equation}\tag{M3}
\log \Lambda_{3}(\bm s)=\log \Lambda_{0}(\bm s)+  \mathrm{Slope}(\bm s)\times  x_{\mathrm{Slope}}^{\mathrm{CAR}}(\bm s).
\end{equation}
By analogy with the latent spatial effect $\bm x_{\mathrm{SU}}^{\mathrm{CAR}}$,  the prior on $\bm x_{\mathrm{Slope}}^{\mathrm{CAR}}$ corresponds to a Gaussian process with CAR structure, and with its own precision parameter $\tau_{\mathrm{Slope}}^{\mathrm{CAR}}>0$, for which we  set an informative prior distribution according to $\mathrm{Pr}(\sqrt{1/\tau_{\mathrm{Slope}}^{\mathrm{CAR}}}>0.1) = 0.01$. %\TC{(Raphael: is that correct?)}

\subsubsection{Model 4: Nonlinear effect and space-varying regression}
In this model, we combine both the nonlinear Slope Steepness effect of M2 and the SVR-coefficient component of M3 into a single model, leading to the following structure:
%Model 26
\begin{equation}\tag{M4}
\log \Lambda_{4}(\bm s)=\log \Lambda_{0,-\mathrm{Slope}}(\bm s)+  x_{\mathrm{Slope}}^{\mathrm{RW1}}(\bm s)+  \mathrm{Slope}(\bm s)\times  x_{\mathrm{Slope}}^{\mathrm{CAR}}(\bm s),
\end{equation}
where hyperparameter priors are fixed as above.

\subsubsection{Model 5: Parsimonious space-varying regression (P-SVR)}

Finally, we construct a model similar to M4 but which links the latent spatial effect $x_{\mathrm{SU}}^{\mathrm{CAR}}(\bm s)$ and the SVR component. If the latent spatial effect acts as a proxy for the precipitation trigger, then its low values indicate a weak or absent trigger effect, and then the Slope Steepness value becomes irrelevant since no landslides occur, whatever the geomorphological conditions.  In this case, the SVR may locally neutralize (i.e., counteract) the globally estimated Slope Steepness effect. We here consider the following parsimonious model:
%(Model 41)
\begin{equation}\tag{M5}
\log \Lambda_{5}(\bm s)=\log \Lambda_{0,-\mathrm{Slope}}(\bm s)+ x_{\mathrm{Slope}}^{\mathrm{RW1}}(\bm s)+  \beta\times \mathrm{Slope}(\bm s)\times  x_{\mathrm{SU}}^{\mathrm{CAR}}(\bm s),
\end{equation}
with the interaction coefficient $\beta\in\mathbb R$ to be estimated. Unlike the more complex model M4, this model features only one single CAR effect, $x_{\mathrm{SU}}^{\mathrm{CAR}}(\bm s)$, instead of the two a priori independent effects, $x_{\mathrm{SU}}^{\mathrm{CAR}}(\bm s)$ and $x_{\mathrm{Slope}}^{\mathrm{CAR}}(\bm s)$, in model M4, such that we consider it as a parsimonious variant of space-varying regression. The prior for the parameter $\beta$ is set to be moderately informative; it is Gaussian with mean $1$ and precision $10$.

\section{Approximate Bayesian inference}\label{sec:inference}
\subsection{The integrated nested Laplace approximation (INLA)}\label{sec:inla}
%\TC{\bf SHOULD WE MAKE A SEPARATE SECTION CONTAINING 3.3 AND 3.4 ONLY?}
INLA \citep{Rue.Martino.Chopin.2009} has found widespread interest in a wide range of applications \citep{Krainski.etal:2018,Lombardo.etal:2018,Moraga:2019,Opitz.etal:2018}, thanks to its ability to provide fast and accurate posterior inference for the general class of latent Gaussian models including log-Gaussian Cox processes \citep{Tierney.Kadane.1986}. The \texttt{R-INLA} package (see \href{http://www.r-inla.org/}{http://www.r-inla.org/}), in which the core statistical methodology is efficiently and conveniently implemented, privileges sparse matrix calculations in large dimensions through systematic use of Gauss--Markov conditional independence structures. With hierarchically structured models including several components with different structures at the latent layer, INLA is typically faster and simpler to tune than simulation-based Markov chain Monte Carlo. (MCMC) methods  \citep{Illian.al.2012,Opitz.2017,Rue.Held.2005,Rue.Martino.Chopin.2009,Rue.al.2016}. 
% for variables such as fixed and random effects, intensities $\Lambda(s_i)\mid \bm X_{\mathcal{S}}$, and model hyperparameterarameter $\tau\mid \bm X_{\mathcal{S}}$ in our conditional autoregressive slope unit model \eqref{eq:spateff}.s, such as the precision p 

In our context of log-Gaussian Cox processes, we write $\Lambda_i=\exp(x_i)$ for the stochastic point process intensity at pixel $i\in\{1,\ldots,n_{\mathrm{grid}}\}$ with $n_{\mathrm{grid}}=449,038$,  where $\bm x=(x_1,\ldots,x_{n_{\mathrm{grid}}})^T$
%,\ldots)$  
denotes the vector of the pixel-based latent Gaussian log-intensities for the whole study area.
We further use the notation $\bm x_{\mathrm{full}} = (\bm x^T, \ldots)^T$ for a vector with $n_{\mathrm{full}}$ components, where ``$\ldots$'' refers to the variables of the additive random effects included in the log-intensity model, e.g., $\bm x^{\mathrm{CAR}}_{\mathrm{SU}}$, and so forth.  
The vector of hyperparameters (i.e., precisions of CAR, RW1 and i.i.d.\ components) is denoted by $\bm{\theta}$ and has $n_\theta$ components. 
The distribution of pixel-based landslide counts $y_i$, collected into a vector $\bm y = (y_1,\ldots,y_{n_{\mathrm{grid}}})^T$, is assumed to be conditionally independent given the (random) intensity values, i.e., 
\begin{equation*}\label{eq:ppdiscr}
y_i \mid \Lambda_i \stackrel{\mathrm{ind.}}{\sim} \mathrm{Poisson}(C\Lambda_i), \qquad i=1,\ldots, n_{\mathrm{grid}},
\end{equation*}
where $C=(15 m)^2$ is a scaling factor corresponding to the area of one pixel.
%(expressed in the unit of $\Lambda$). 
The principal inference goal is the calculation of the posterior densities of hyperparameters and of the components of $\bm x_{\mathrm{full}}$, the latent vector with multivariate Gaussian prior distribution, i.e.,
\begin{align}
\pi(\theta_j \mid \bm y) &= \int \pi(\bm x_{\mathrm{full}}, \bm \theta \mid \bm y)
\,\mathrm{d}\bm x_{\mathrm{full}}\,\mathrm{d}\bm \theta_{-j}, \label{eq:posttau}, \quad j=1,\ldots, n_\theta,\\
\pi(x_i\mid \bm y) &= %\int\int \pi(\bm x, \bm \theta\mid \bm y) \mathrm{d}\bm x_{-i}\, \mathrm{d}\bm\theta =  
\int \pi(x_i \mid \bm\theta, \bm y) \pi(\bm \theta\mid \bm y)\,\mathrm{d}\bm\theta, \quad i=1,\ldots,n_{\mathrm{full}}. \label{eq:postlambda}
\end{align}
However, calculation of \eqref{eq:posttau} and \eqref{eq:postlambda} is hampered by the high-dimensional numerical integration over the space $\mathbb{R}^{n_{\mathrm{full}}}$ spanned by the Gaussian vector $\bm x_{\mathrm{full}}$.  Instead, INLA  uses the Laplace approximation, which corresponds to replacing integrand functions  by suitable Gaussian density approximations.  On the other hand, the  integration with respect to the components of $\bm{\theta}$ is done through numerical integration schemes, such that only a small number of hyperparameters can be estimated. %\TC{(Raphael: equations should only be numbered if it is necessary to refer to them... we should check that for each equation)}

\subsection{Model comparison and selection}\label{sec:modelcomparison}
First, we propose to compare models through the classical information criteria DIC and WAIC. These goodness-of-fit criteria take the effective dimension of the latent model into account, thus penalizing model complexity. Their close relationship to the predictive performance measured through leave-one-out cross-validation has been established, and WAIC is known to better take the stochasticity of the posterior predictive distributions into account \citep{Gelman.al.2014}. With INLA, these quantities are calculated through sensible approximation techniques \citep{Rue.Martino.Chopin.2009}.

To focus more directly on criteria evaluating the spatial predictive performance, we also devise a $10$-fold cross-validation scheme.  Specifically, we randomly partition the SUs into 10 folds, each containing (approximately) the same number of SUs. We calculate predictive scores for two mapping units: pixels and SUs, for the latter by aggregating observed and predicted counts over the pixels of each SU. 
At the pixel level, we  consider predictive scores that use either the predicted counts $\hat{\lambda}_i = \mathbb{E}(\Lambda_i\mid \bm y)$, the predicted probabilities of within-pixel landslide occurrences $\hat{p}_i=1-\exp(-\hat{\lambda}_i )$, or the full posterior predictive distribution of ${\Lambda_i\mid \bm y}$. Similarly, at the SU level, we consider the predicted counts estimated as $\hat{\lambda}_{\mathrm{SU}}= \sum_{i\in\mathrm{SU}}\hat{\lambda}_i$, the predicted probabilities of within-SU landslide occurrences $\hat{p}_{\mathrm{SU}} = 1-\exp(-\hat{\lambda}_{\mathrm{SU}} )$, and the posterior predictive distributions of $\left(\sum_{i\in\mathrm{SU}}\Lambda_i\right)\mid \bm y$. An alternative approach for predictive diagnostics, studied by \citet{Leininger.Gelfand.2017}, would be to construct hold-out sets by removing  points at random from the point pattern; this is known as \emph{thinning}. Here, we prefer the more challenging task of predicting entire spatially-contiguous areas where all data within SUs have been removed, which is also more suited to assessing slope-wise landslide hazard. 

The posterior predictive distributions are obtained by generating a large number of posterior samples of counts for each cross-validation fit. While INLA does not directly provide posterior samples because of its use of analytical and not simulation-based approximations, these can be generated conveniently \citep{Rue.al.2016}. Using \texttt{R-INLA}'s internal, discrete approximations for posterior distributions of hyperparameters and latent Gaussian fields,  the simulation algorithm first generates a realization of the hyperparameter vector; next,  conditional on these hyperparameters, a latent Gaussian field is sampled; finally, counts are simulated from  the pixel-based Poisson distributions with intensities defined according to the simulated latent Gaussian field. In what follows, cross-validation results using simulations of the posterior predictive distributions are based on $5000$ samples of the full posterior model.

We consider four types of cross-validated predictive scores: the area-under-the-curve (AUC) \citep{Fawcett.2006} to measure prediction quality for the presence or absence of landslides within mapping units; the residual sum of squared errors (RSS) and the residual sum of absolute errors (RSA), both using predicted and observed counts; and the continuous ranked probability score \citep[CRPS,][]{Gneiting.Katzfuss.2014} using the predictive distribution functions and observed counts. The formulas for pixel-based RSS and RSA are as follows:  
$$
\mathrm{RSS}_{\mathrm{grid}} = \sum_{i=1}^{n_{\mathrm{grid}}} (y_i-\hat{\lambda}_i)^2,\quad \mathrm{RSA}_{\mathrm{grid}} = \sum_{i=1}^{n_{\mathrm{grid}}} |y_i-\hat{\lambda}_i|,
$$
where $\hat{\lambda}_i=\int_0^\infty \pi_i\{\log(\lambda)\mid \bm y\}\, \mathrm{d}\lambda$ %\TC{(Raphael: shouldn't there be a Jacobian term $1/\lambda$ here? the notation is not super clear to me...)} 
with $\pi_i\{\cdot \mid \bm y\}$  the posterior density of $x_i=\log (\Lambda_i)$. The general CRPS formula for a single observation $y_{\mathrm{obs}}$ and a corresponding (posterior) predictive distribution $\hat{F}(y)$ from a model may be expressed as $\int_{-\infty}^\infty {\{\hat{F}(y)-\mathbf{1}(y\geq y_{\mathrm{obs}})\}^2}\,\mathrm{d}y$. For the pixel-based CRPS in our case, we add up the CRPS values over all pixels and therefore use
$$
\mathrm{CRPS}_{\mathrm{grid}}=\sum_{i=1}^{n_{\mathrm{grid}}} \sum_{y=0}^{\infty}\sum_{k=0}^{y}\left[ \int_0^\infty   \exp(-\lambda) \frac{\lambda^{k}}{k!} \pi_i\{\log(\lambda)\mid \bm y\}\frac{1}{\lambda}\, \mathrm{d}\lambda - \mathbf{1}(y \geq y_i)\right]^2. 
$$
Analogous formulas are used for SU-based criteria, where pixel-based observed counts $y_i$ and intensities $\lambda_i$ must be aggregated over the pixels for each SU. This requires resorting to the joint posterior distributions of all $x_i$ corresponding to the pixels in a given SU. Since such CRPS formulas are difficult to calculate analytically, we use  posterior sampling as implemented in \texttt{R-INLA} and compute a Monte-Carlo approximation of  CRPS values based on  a large number of posterior samples ($5000$ as above). 

The AUC considers only presence-absence data, which is a strong simplification for assessing the prediction of landslide counts, especially at the SU level where counts larger than one are frequent. By contrast, the other three measures rely on counts: while RSS and RSA focus on point predictions defined through  the posterior mean of intensities at pixel level, CRPS also accounts for the uncertainty of the predictive distributions and yields good scores for models that provide predictions that are both calibrated (i.e., correct on average) and sharp (i.e., having little prediction uncertainty). We calculate these four predictive scores for each of the 10 folds, both at pixel and SU levels,  and finally we average the predictive scores of the 10 folds together. Lower final values correspond to better predictive performances.

\section{Results and discussion}
\label{sec:results}

\subsection{Comparison of models using basic diagnostics} 
%\TC{\bf FIND A BETTER TITLE FOR THIS SUBSECTION? Raphael: I think the title is OK...}
Table~\ref{tab:modcomp}  reports results for our models fitted to the full dataset, including the precision parameter (i.e., inverse variance) of the estimated latent spatial effect (LSE) $\bm x_{\mathrm{SU}}^{\mathrm{CAR}}$ included in all models, the effective number of parameters (i.e., the effective dimension of the linear predictor when accounting for the dependence between the latent variables), and the two information criteria DIC and WAIC. 
%We point out that $n_{\mathrm{eff}}$, DIC, and WAIC are only approximate ``point estimates" of quantities which are defined theoretically  as expectation values, such that relatively small differences in such values should not be overinterpreted. 
The similar LSE precisions indicate that the variability of the LSE is relatively stable over different models, even with the most complex models. The effective number of parameters is relatively similar for all models except the model with an i.i.d.\ effect resolved at the pixel scale; recall that there is a large number $449,038$ of pixels. DIC and WAIC values are relatively similar overall, although both information criteria give a clear preference to models with space-varying regression components. While DIC ranks first the most complex model M4 with independently specified RW1- and SVR-components for the Slope Steepness, WAIC prefers  model  M3, which includes only a fixed (i.e., global and linear) Slope Steepness effect and the SVR component. 

We now also report and discuss estimated precision parameters for the specific components added in the models extending the baseline.  In M1a with pixel-resolved i.i.d.\ effect, we estimate a posterior precision of about $1.5$ for the i.i.d\ component, which indicates the presence of a  rather strong independent effect at the pixel scale, not explained by the aggregated view based on SUs, and without dependence spanning over neighboring pixels. By contrast, the precision of the SU-resolved i.i.d.\ effect in M1b is very high (180), indicating a relatively small contribution of this effect to the model. 
%\TC{(Raphael: please check, I have interchanged M1a and M1b, as the former refers to the pixel-based model, while the latter to the SU-based model as explained above...)}
In M2, the precision of the nonlinear RW1-effect of Slope Steepness is $4.4$. In M3 with an SVR component, the precision of the space-varying coefficient  is $4.3$.   By jointly including the RW1- and SVR-effects of Slope Steepness in M4, we get RW1-precision of $5.8$ and SVR precision of $2.4$.  The former is higher than without the SVR component (M2), indicating that the influence of Slope Steepness is now partially captured by the additional SVR component, whose precision is relatively low. We conjecture that this low precision shows that the SVR component captures the influence of Slope Steepness more easily than the global RW1-effect of Slope Steepness; % \TC{(Raphael: Not sure to understand the argument...)}; 
moreover, the non-continuous specification of the RW1-curve may require stronger variation of the SVR-component's contribution  at relatively small spatial scales to smooth the RW1-effect. Finally, model M5 with a parsimonious SVR component has a RW1-precision estimated of $7.9$, and the posterior mean of the $\beta$-coefficient is given by $0.13$ with credible interval $[0.10,0.16]$. Therefore, a significant transfer of predictive information has taken place from the LSE (whose precision is similar to the other models; recall Table~\ref{tab:modcomp}) to the space-varying Slope Steepness influence, while the RW1-contribution has been reduced compared to the other models with RW1-component. The parsimonious constraint linking the SVR to the latent spatial effect in M5 leads to an improved goodness-of-fit compared to the baseline M0 and the models with additional i.i.d.\ components, but based on its DIC and WAIC values we conclude that it  cannot fully attain the high flexibility of models M3 and M4. 

\begin{table}
	\centering
	\caption{Comparison of fitted models in terms of the estimated precision of the latent spatial effect (LSE) $\bm x_{\mathrm{SU}}^{\mathrm{CAR}}$, and information criteria (DIC and WAIC, respectively). The column $n_{\mathrm{eff}}$ denotes the effective number of parameters of the fitted model. %\TC{(Raphael: I interchanged the rows corresponding to M1a and M1b, as M1a refers to the pixel-iid model and M2b to the SU-iid model, as described above... please double-check that it's correct.)}
	}\label{tab:modcomp}
	\begin{tabular}{l|rrrr}
		Model & LSE precision & $n_{\mathrm{eff}}$ & DIC & WAIC\\
		\hline
M0 (baseline) & 0.35 (0.06) & 1006 & 40142 & 39869\\ 
M1a (pixel-iid) & 0.3 (0.02) & 4166 & 41925 & 40834\\ 
M1b (SU-iid) & 0.32 (0.02) & 987 & 40120 & 39867\\ 
M2 (RW1) & 0.31 (0.02) & 1002 & 40132 & 39864\\ 
M3 (SVR) & 0.31 (0.02) & 1144 & 40005 & {\bf 39690}\\ 
M4 (RW1-SVR) & 0.31 (0.01) & 1203 & {\bf 39949} & 39745\\ 
M5 (P-SVR) & 0.34 (0.02) & 1016 & 40087 & 39817\\ 
	\end{tabular}
\end{table}

\subsection{Influence of fixed effects}
\begin{figure}[t!]
	\centering
		\includegraphics[width=0.7\linewidth]{\FIGS 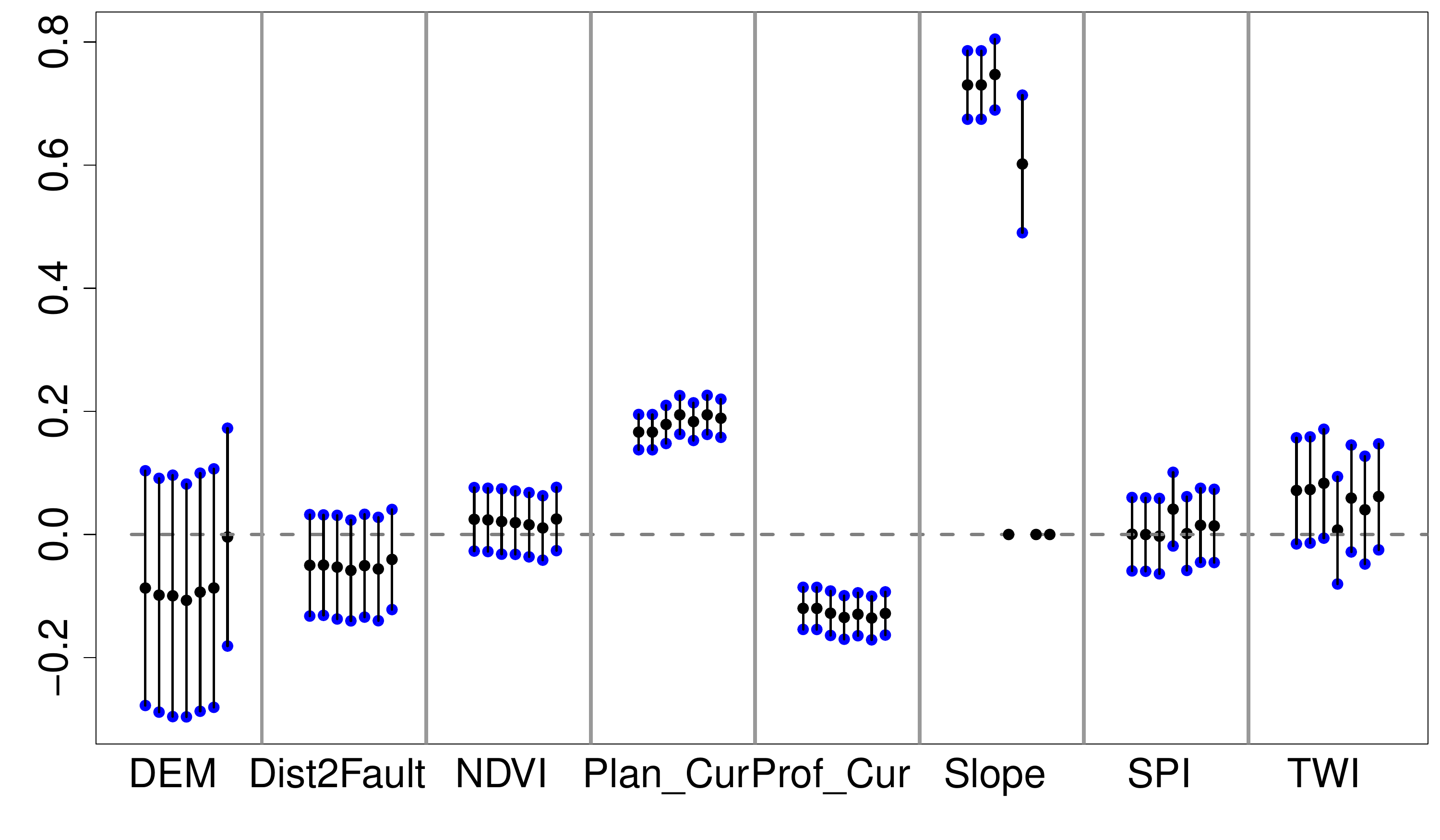} 
	\caption{Estimated fixed effect coefficients (except the intercept) for the 7 models (M0, M1a, M1b, M2, M3, M4, M5, from left to right in each panel corresponding to one of the covariates). Black dots show posterior means, while vertical segments and blue endpoints indicate the size of $95\%$ credible intervals. Fixed effect coefficients for Slope Steepness are fixed to $0$ for some models and appear only through a black dot at level $0$  in these cases.}
	\label{fig:fixed}
\end{figure}

In Figure~\ref{fig:fixed}, we compare the estimated coefficients for the 8 predictor variables included as fixed effects in our different models. The contribution of these covariates, and the associated uncertainty of their coefficients, are estimated to be very similar across the 7 models studied here, except for the Slope Steepness due to the major differences in model structure with respect to the contribution of this covariate. Interestingly, the parsimonious SVR structure seems to have fully absorbed the influence of DEM into the Slope-SVR part of the model, $\mathrm{Slope}(\bm s)\times x_{\mathrm{Slope}}^{\mathrm{CAR}}(\bm s)$. We do not show results for the categorical covariates, but the conclusions remain qualitatively similar. 

\begin{figure}[t!]
	\centering
	\includegraphics[width=.32\linewidth]{\FIGS 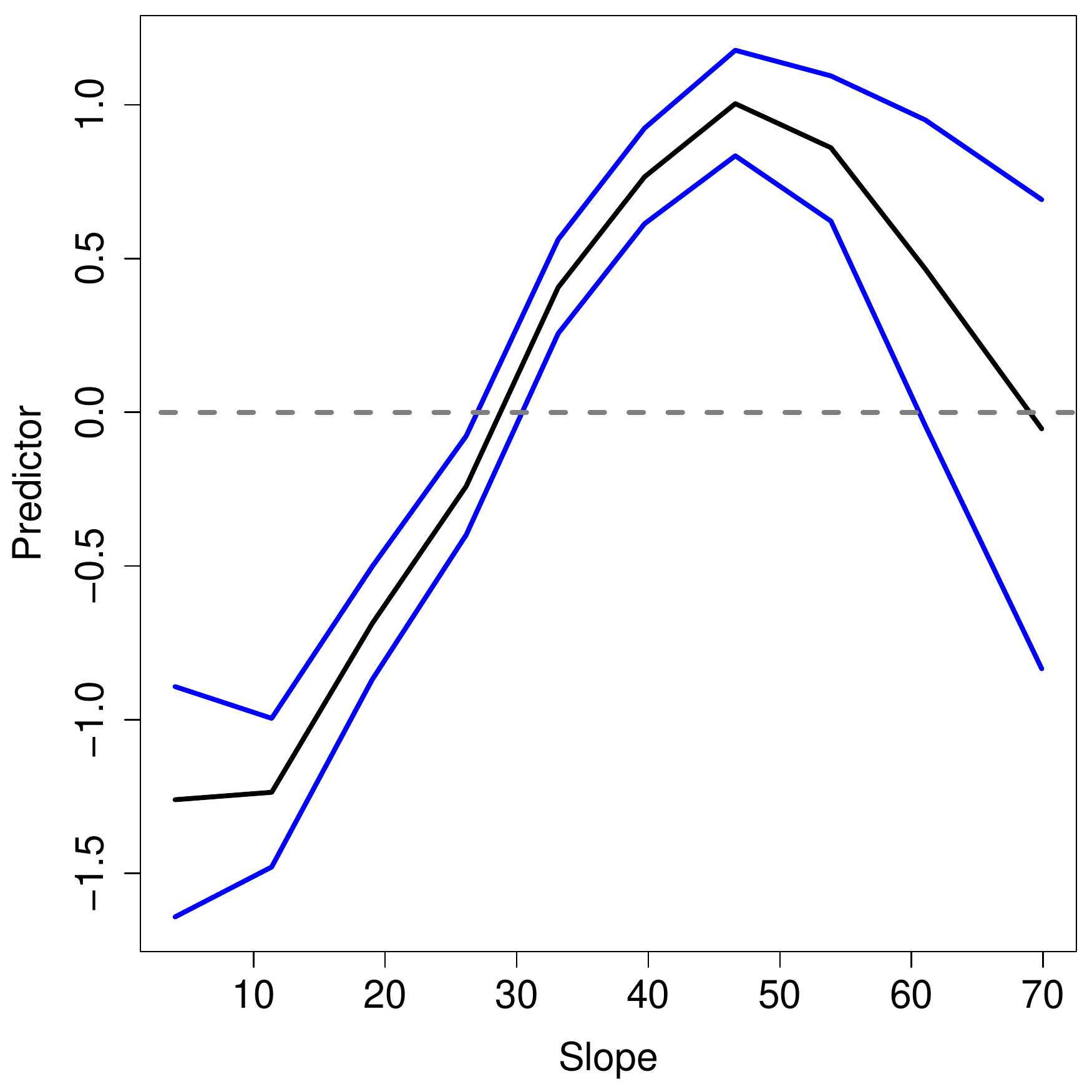} 
	\includegraphics[width=.32\linewidth]{\FIGS 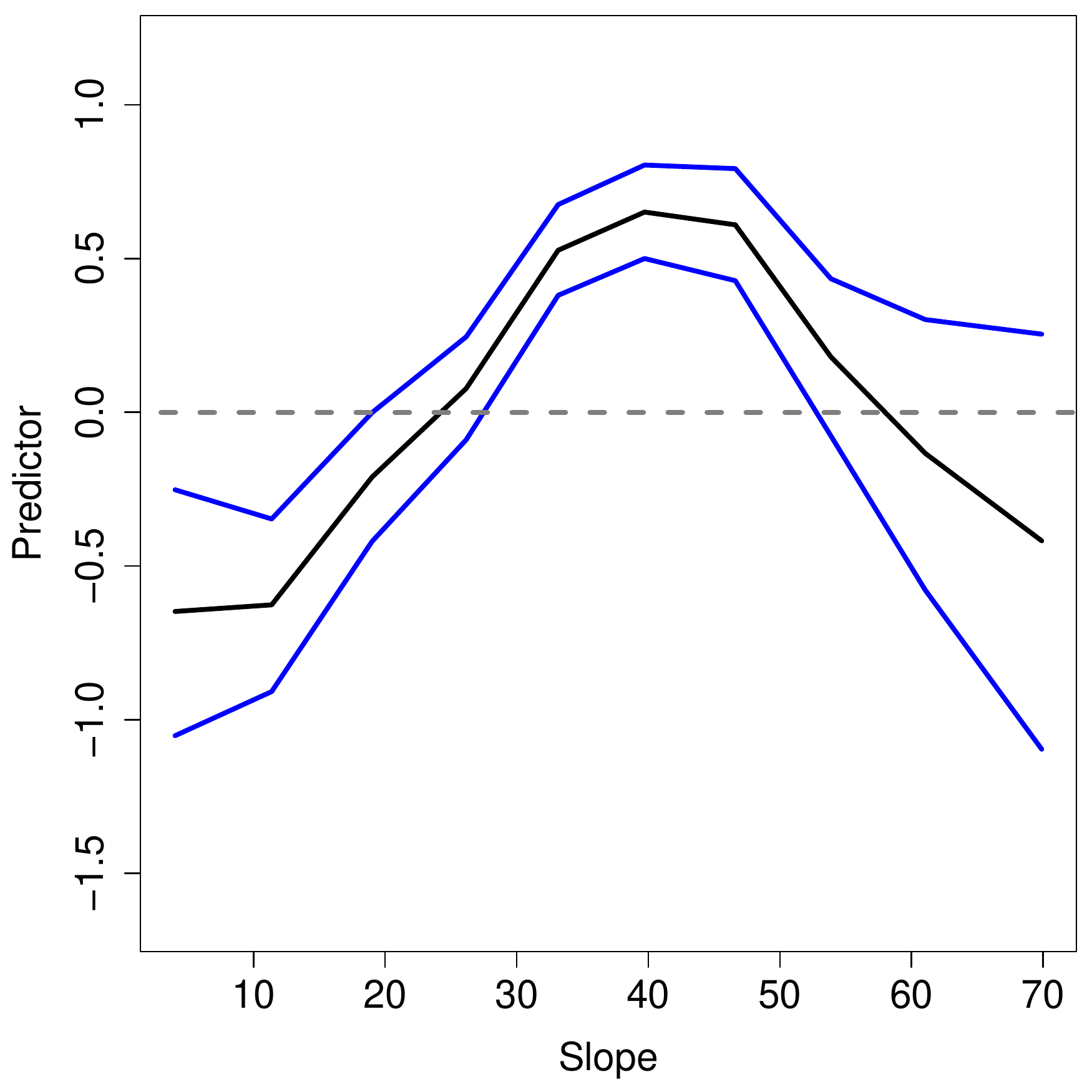} 
	\includegraphics[width=.32\linewidth]{\FIGS 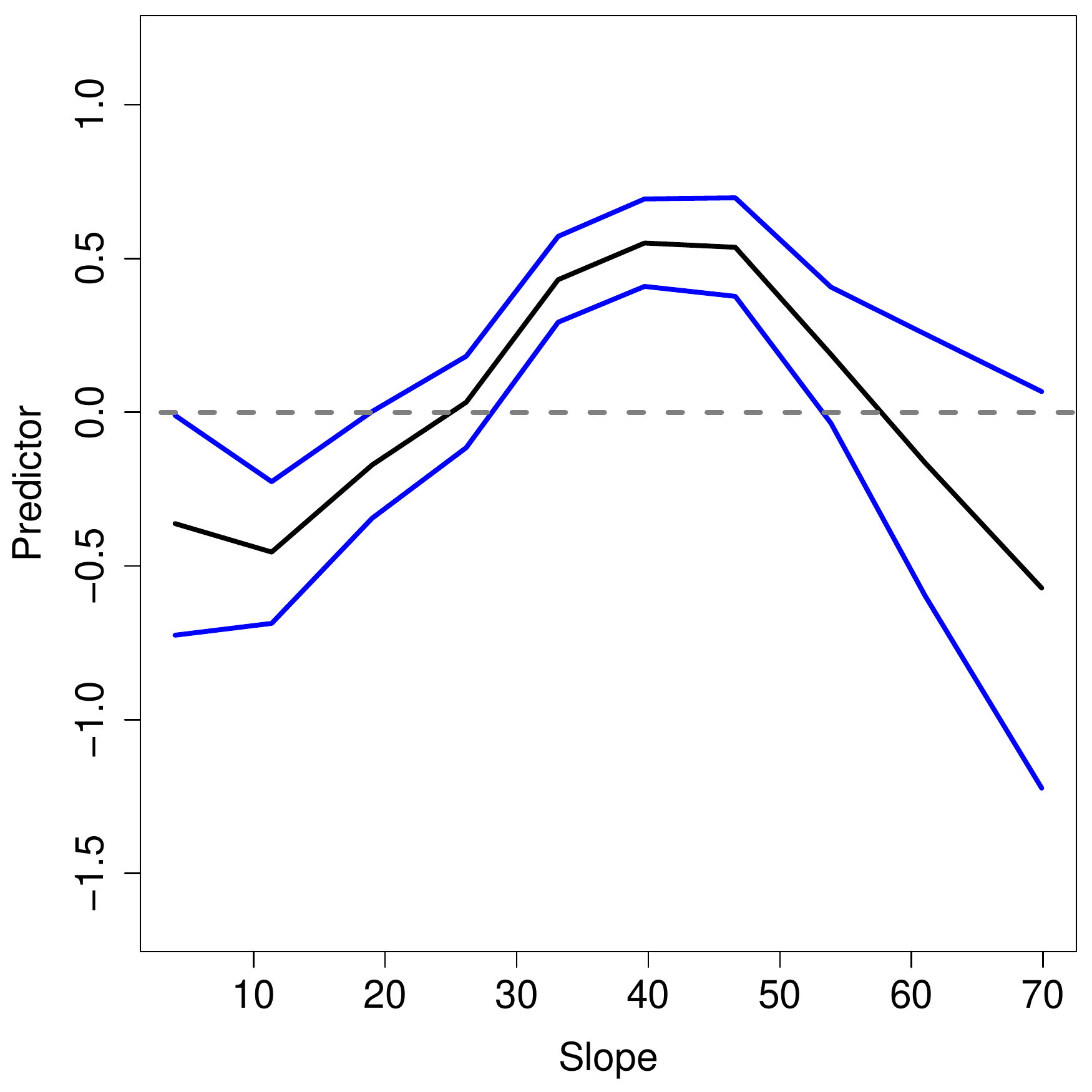} 
	
	\caption{Estimated RW1-effects for the Slope Steepness (from left to right, for models M2, M4 and M5). Posterior means are shown in black, $95\%$ pointwise credible intervals in blue.}  %Aspect (M0, left display) 
	\label{fig:sloperw1}
\end{figure}

For the models M2, M4 and M5 with a nonlinear Slope Steepness effect modeled through a RW1 component, Figure~\ref{fig:sloperw1} shows the resulting estimated curves---for better readability of the plots, piecewise constant curves are replaced by piecewise continuous interpolations. Nonlinear influence is obvious from these plots and displays a similar bell shape in all three models, with intermediately steep slopes between 30 and 60 degrees presenting high relative risk of landslide occurrence. Models M4 and M5 include additional SVR-components to capture the slope-specific influence at SU resolution, such that the estimated RW1 curves appear to be flatter, which implies smaller variations in relative risk. 

\subsection{Spatial predictive  performance comparison}
%cross-validation to check predictive performance through out-of-sample prediction: fast thanks to INLA
%criteria: AUC (but only binary), CRPS, ...
%different spatial units

Table~\ref{tab:cv} reports spatial predictive scores based on a 10-fold stratified cross-validation, where folds are composed of a random selection of entire SUs. For all models,  pixel- and SU-based AUC values are very close to each other and reach around $0.9$, indicating a very good performance for predicting the presence of landslides, especially at the SU level. When considering sums of absolute and squared errors (RSA and RSS, respectively), stronger differences between models arise, with model M2 with nonlinear Slope influence obtaining the best scores, and M1a with pixel-based  i.i.d.\ effects performing much worse than the other models.  
%\TC{(Raphael: I replaced M1b by M1a to be consistent with the definition above...)}. 
Throughout, model M2 has a very good performance and has the best score except for the SU-based AUC value, although differences are rather small. 
%and potentially non-significant). 
Therefore, the inclusion of a nonlinear effect of the Slope Steepness, here implemented through a random effect with RW1 prior, is important for good prediction. The baseline model shows stable and  good  performance throughout and does not suffer from some relatively bad count-based scores arising for some of the extended models (except M2).  Overall, the ranking of models based on their predictive performance looks quite different from the one based on goodness-of-fit measured through information criteria in Table~\ref{tab:modcomp}. A possible reason is that very high stochasticity and complexity of prior models may lead to more unstable, noisy predictions. We recommend a careful inspection of the fitted models based on several criteria, for goodness-of-fit and for out-of-sample prediction. The ``best'' model M2 is more complex than the baseline model M0, but we add only a relatively small number of $10$ latent components to achieve a nonlinear contribution of Slope Steepness. We also stress that, for our landslides data, the inclusion of i.i.d.\ effects (pixel- or SU-based) could not provide substantial improvements of goodness-of-fit or predictive performance. 
Finally, the models M3, M4 and M5, which possess extra flexibility thanks to a space-varying regression component, are relatively competitive overall, despite their relatively worse performance on RSS measures. Such models can still be useful by offering insights into the ``physical'' interaction of Slope Steepness with the unobserved precipitation trigger, as further explained in Section~\ref{sec:interpretation} below.
%\TC{(Raphael: I feel we could discuss a little bit more the performance of SVR models M3, M4, M5 here... Is the extra flexibility worth the effort? Perhaps we could argue that despite having a worse predictive performance, they can still be useful by offering insights into the ``physical'' interaction of Slope Steepness with the unobserved precipitation trigger, as explained in Section 5.5 below...)}

\begin{table}
	\centering
	\caption{Cross-validation-based comparison of spatial predictive performance of fitted models, with the score of the best-performing models shown in bold face. Scores are given with 4 significant digits. Mathematical details on the different scores are given in Section~\ref{sec:modelcomparison}. Pixel-based and SU-based scores are denoted with the subscripts $_{\mathrm{grid}}$ and $_{\mathrm{SU}}$, respectively. 
%\TC{(Raphael: I interchanged the rows corresponding to M1a and M1b, as M1a refers to the pixel-iid model and M2b to the SU-iid model, as described above... please double-check that it's correct.)}
    }\label{tab:cv}
	\begin{tabular}{l|rrrrrrrr}
		Model & AUC$_{\mathrm{grid}}$ & AUC$_{\mathrm{SU}}$ & RSA$_{\mathrm{grid}}$   &  RSA$_{\mathrm{SU}}$ & RSS$_{\mathrm{grid}}$ & RSS$_{\mathrm{SU}}$ & CRPS$_{\mathrm{grid}}$ & CRPS$_{\mathrm{SU}}$ \\
		\hline
M0 (baseline) & 0.8958 & 0.9308 & 420.0 & 420.5 & 2614 & 2608 & 466.3 & 240.6\\ 
M1a (pixel-iid) & 0.8960 & 0.9308 & 590.1 & 590.0 & 5628 & 5602 & 474.7 & 279.7\\
M1b (SU-iid) & 0.8958 & 0.9309 & 420.6 & 421.2 & 2615 & 2635 & 466.3 & 240.5\\  
M2 (RW1) & {\bf 0.8956} & 0.9310 & {\bf 411.3} & {\bf 411.5} & {\bf 2481} & {\bf 2483} & {\bf 464.7} & {\bf 238.5}\\ 
M3 (SVR) & 0.8963 & 0.9305 & 441.1 & 441.7 & 2922 & 2936 & 467.9 & 242.8\\ 
M4 (RW1-SVR) & 0.8964 & {\bf 0.9302} & 436.6 & 437 & 2770 & 2801 & 465.9 & 241.1\\ 
M5 (P-SVR) & 0.8966 & 0.9311 & 430.1 & 429.8 & 2833 & 2819 & 466.4 & 241.3\\ 
	\end{tabular}
\end{table}

\subsection{Landslide intensity mapping}

\begin{figure}[t!]
	\centering
	\begin{tabular}{cc}
		\includegraphics[width=6cm]{\FIGS 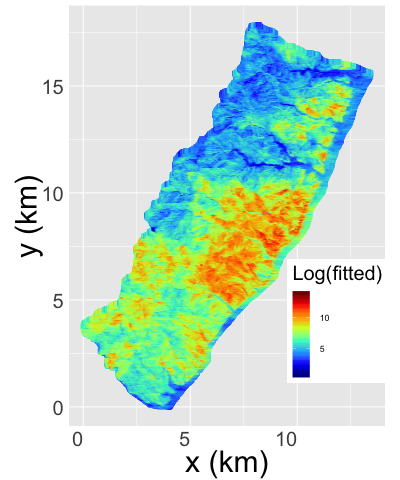} & 
		\includegraphics[width=6cm]{\FIGS 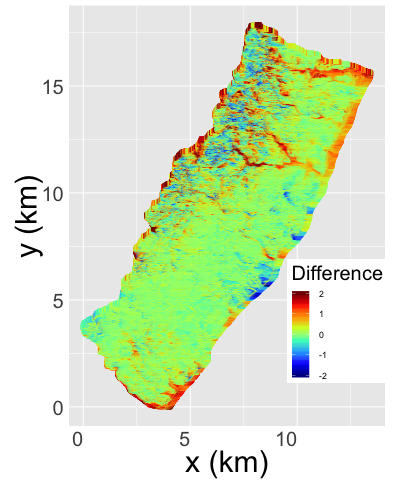}
	\end{tabular}
	\caption{Posterior mean of the estimated log-intensity at the pixel scale. Left: baseline model M0. Right: Difference of log-intensity between models M4 (RW1-SVR) and M0. For better visualization, a small number of difference values outside the interval $[-2,2]$ have been replaced by $-2$ (if value $<-2$) or by $2$ (if value $>2$).}
	\label{fig:intensity}
\end{figure}

In Figure~\ref{fig:intensity}, we show the posterior mean of the estimated log-intensity (at pixel scale) of model M0, and the difference in log-intensity between the most complex model Model M4 (RW1-SVR) and M0. In the log-intensity of M0, the influence of geological structures such as river valleys (very low intensity) and mountain ridges comes out clearly. The spatial structure of the predicted values is dominated by the spatial effect $x_{\mathrm{SU}}^{\mathrm{CAR}}(\bm s)$ capturing the influence of the spatial variation of the precipitation trigger. In this model, the amplitude of log-predicted intensities is close to that of the posterior mean of the spatial effect, which is evidence that the spatial effect is crucially necessary to account for unobserved covariate effects and to locally counteract the influence of observed covariates. The latter may be locally mis-estimated depending on the force of the precipitation trigger. Differences between models M0 and M4 are usually relatively minor, but in some small sub-areas, especially river valleys, model M4 has substantially higher log-intensity values. 

\subsection{Interpretation of the Slope Steepness contribution}
\label{sec:interpretation}

%Role of the precipitation trigger
\begin{figure}[t!]
	\centering
	\begin{tabular}{cc}
		\includegraphics[width=5.5cm]{\FIGS 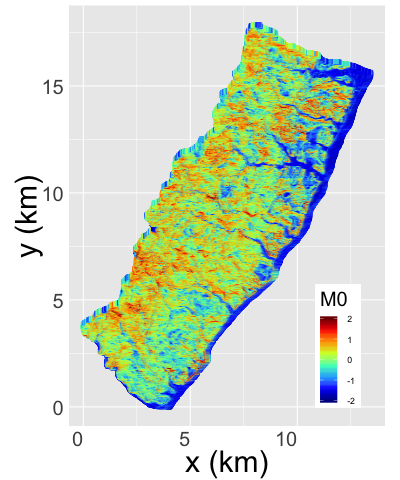} & 
		\includegraphics[width=5.5cm]{\FIGS 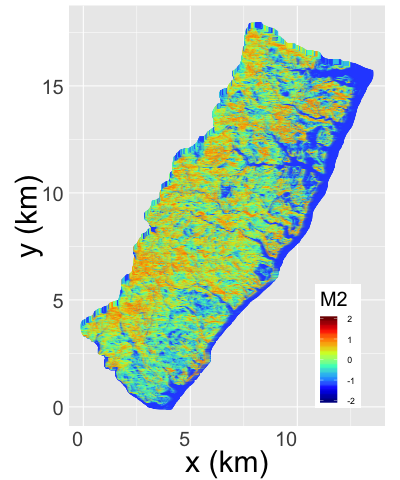} \\
		\includegraphics[width=5.5cm]{\FIGS 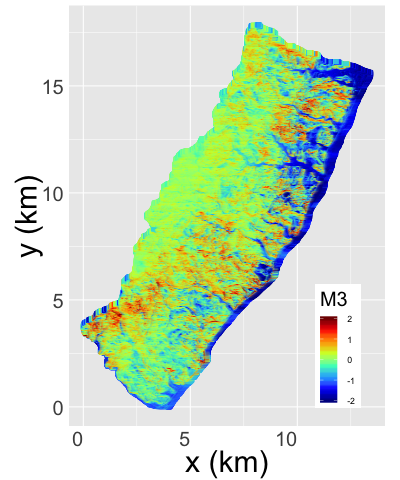} &
		\includegraphics[width=5.5cm]{\FIGS 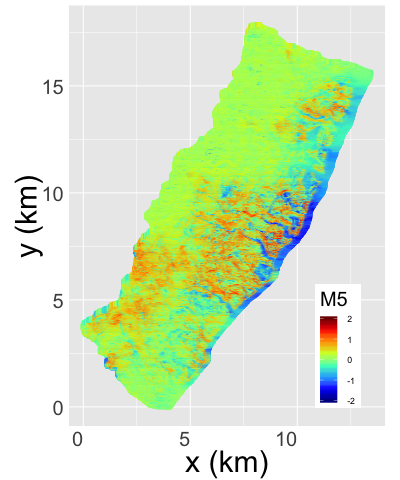}
	\end{tabular}
	\caption{Contribution of the Slope Steepness covariate to the linear predictor (here expressed through the posterior mean) according to the following models: M0 (fixed effect; upper left); M2 (RW1 effect; upper right); M3 (SVR; bottom left); M5 (P-SVR; bottom right). For better visualization, very small number of values outside the interval $[-2,2]$ have been replaced by $-2$ (if value $<-2$) or by $2$ (if value $>2$).}
	\label{fig:slopeeffect}
\end{figure}

The maps in Figure~\ref{fig:slopeeffect} show how the observed Slope Steepness variable contributes to the linear predictor according to the different model structures (M0, M2, M3, M5).  The SVR models M3 and M5 lead to an overall contribution of Slope Steepness that is strongly conditioned on the precipitation trigger; weakly impacted areas  such as the northwestern part of the study area have a very weak contribution close to 0. In contrast, the simpler models M0 and M2 including a fixed effect or a RW1-effect show spatial variation that is closer to the one of the Slope Steepness.

From a physical perspective, we expect to observe at most a few landslides in areas with low or absent precipitation trigger, and, intuitively, the effect of covariates may therefore become irrelevant in such areas.  %\BakkaComment{I do not agree completely, they might be equally relevant in few-landslides areas, as they are a relative indicator. Clarify?} 
While basic models such as M0 and M2 are unable to capture this behavior, more complex models with space-varying covariate coefficients can adequately reflect such natural physical constraints, and help to better highlight areas where a covariate substantially increases landslide hazard. Our model M5 (parsimonious SVR) is even \emph{designed} to explicitly integrate this interaction between the precipitation trigger (represented here through the latent spatial effect (LSE), $x_{\mathrm{SU}}^{\mathrm{CAR}}(\bm s)$) and the space-varying coefficient. The estimated coefficient $\hat{\beta}=0.13$ in model M5 measures the strength of this interaction, which turns out to be highly significant in our model. 

At sites where the precipitation trigger is present, a strong response to Slope Steepness is expected for intermediate angles. 
%starting at 21 degrees approximately, 
In our models, the LSE acts as a proxy for the influence of the precipitation trigger, and we therefore expect a strong interaction between the precipitation trigger and the Slope Steepness effect in the log-intensity of the point process. In Model M4, the LSE and the space-varying Slope Steepness effect are components without any prior dependence, but we can investigate the posterior model structure, and in particular posterior correlation between these components. In Model M5, the interaction structure is fixed to a linear rescaling determined by the $\beta$-hyperparameter. The plots on the left-hand side of Figure~\ref{fig:slopeeffect} show a smoothed image of the space-varying Slope Steepness coefficient, plotted with respect to a two-dimensional coordinate system given by the estimated posterior mean of the LSE in each SU, and the Slope Steepness value averaged over the pixels of each SU. The original, non-smoothed values were obtained at the points shown as small grey dots. A clear interaction pattern between the LSE and the SVR coefficient arises in M4, with lower LSE corresponding to lower SVR coefficient. We underline that the parsimonious model M5 is able to reproduce a very similar structure. The strong similarity of the results for the two models M4 and M5, despite M5 offering much less flexibility due to its rigid link between SVR-coefficient and LSE, persists in the plots on the right-hand side of Figure~\ref{fig:slopeeffect}. They show the actual space-varying Slope Steepness effect (i.e., $\mathrm{Slope}(\bm s)\times x_{\mathrm{Slope}}^{\mathrm{CAR}}(\bm s)$), which corrects the global RW1-Slope Steepness effect. As expected, the plots show a correction towards lower relative landslide hazard for the combinations of high LSE/flat slopes and small LSE/steep slopes. A relatively strong correction towards higher relative landslide hazard is necessary for high LSE/intermediate slopes, and relatively low LSE but Slope Steepness close to the pivotal value of 20 degrees (or slightly above).
%\TC{(Raphael: cite a paper for the $21^\circ$...)}. 
We conclude that the parsimonious model M5 offers a correction of the trigger-independent RW1-Slope Steepness effect that reflects known physical behavior, and the typical landsliding response behavior to Slope Steepness according to the (approximate) intervals $0$--$20$ degrees, $20$--$35$ degrees and $>35$ degrees. 
Field and empirical evidence suggests that the first interval $0$--$20$ degrees corresponds to slopes that are too flat for landslides to occur, even in the presence of a strong trigger event, while the last interval of $>35$ degrees corresponds to slopes where, typically, material that is susceptible to sliding has already gone in the past. Therefore, the trigger primarily acts on slopes in the interval of $20$--$35$ degrees where landslides occur most easily.  
	
\begin{figure}[t!]
	\centering
	\begin{tabular}{cc}
		\includegraphics[width=5cm]{\FIGS 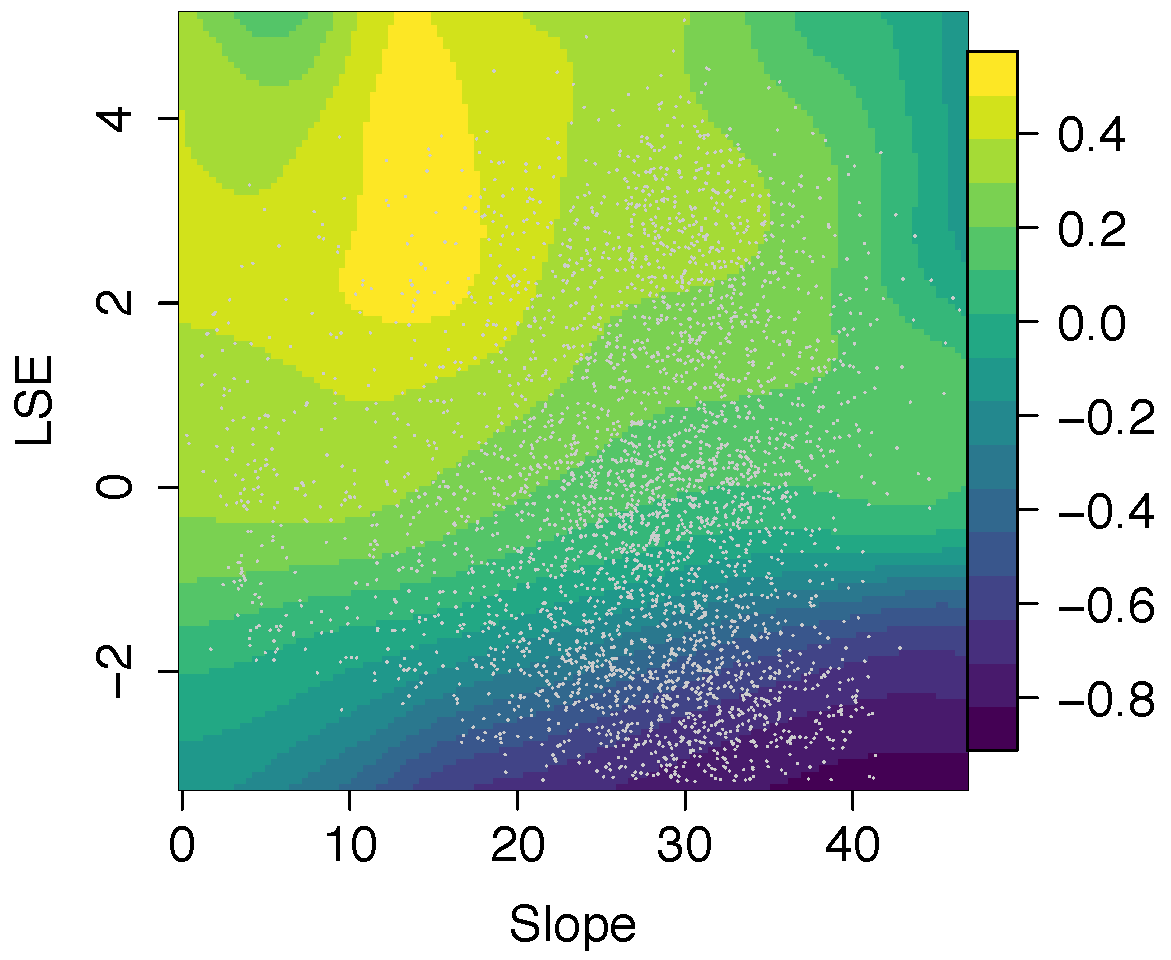} & 
		\includegraphics[width=5cm]{\FIGS 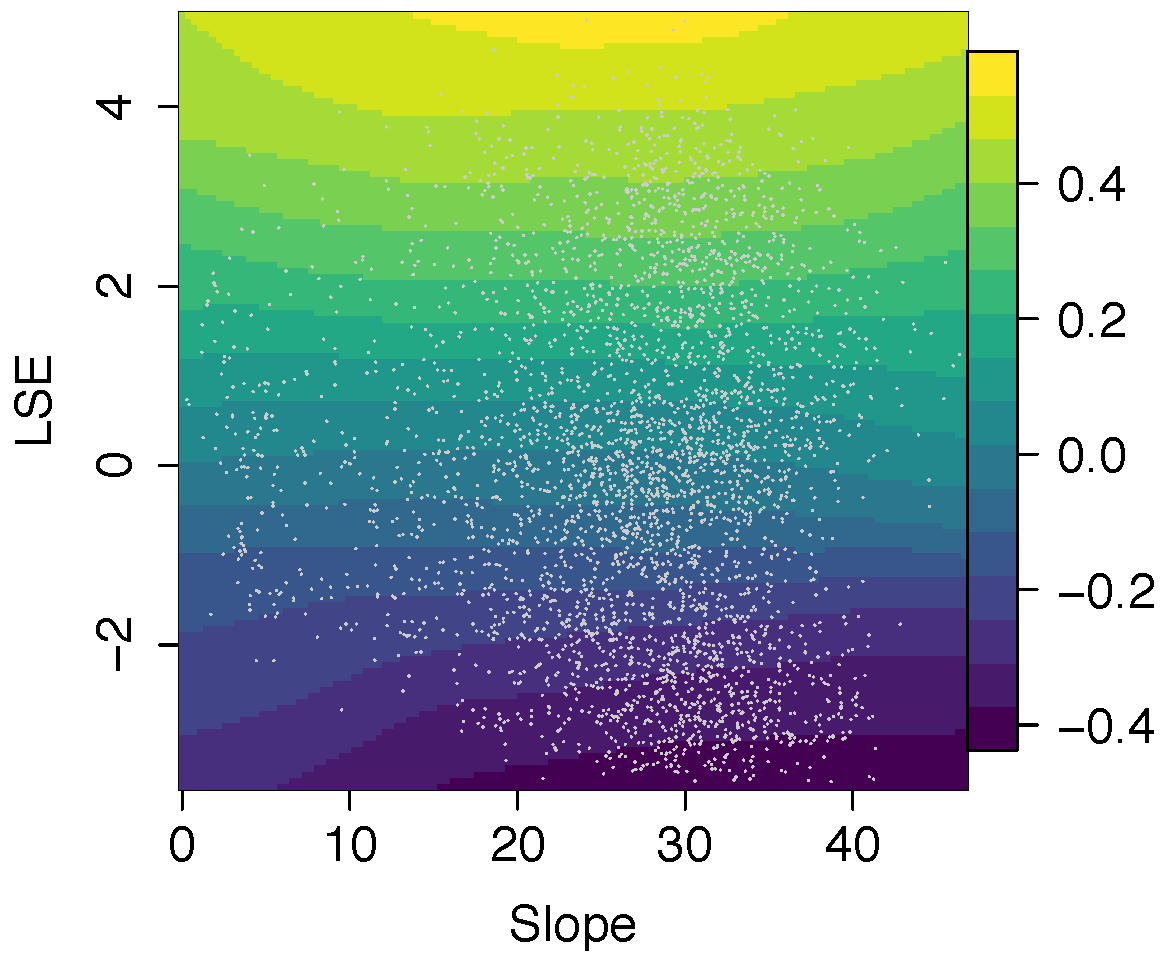} \\
		\includegraphics[width=5cm]{\FIGS 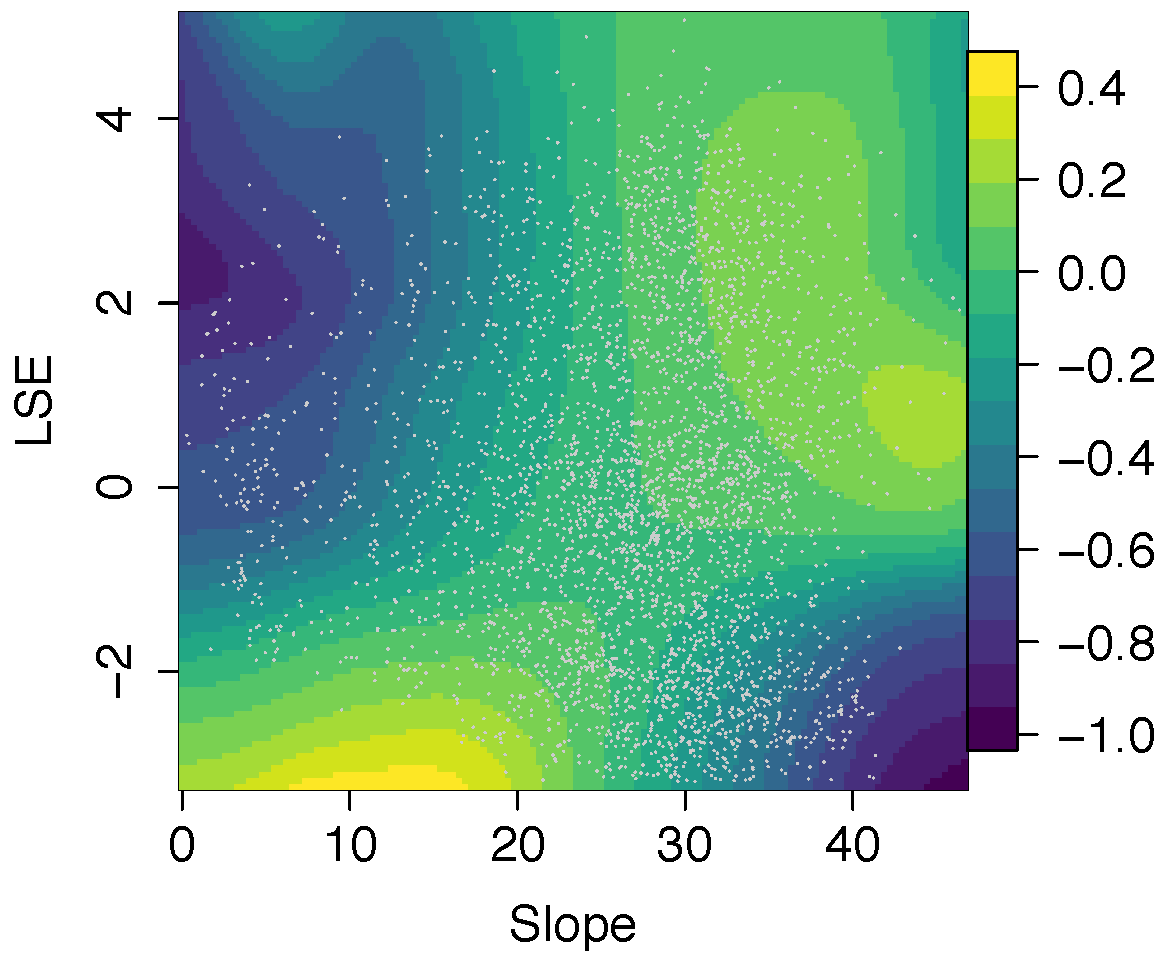} & 
		\includegraphics[width=5cm]{\FIGS 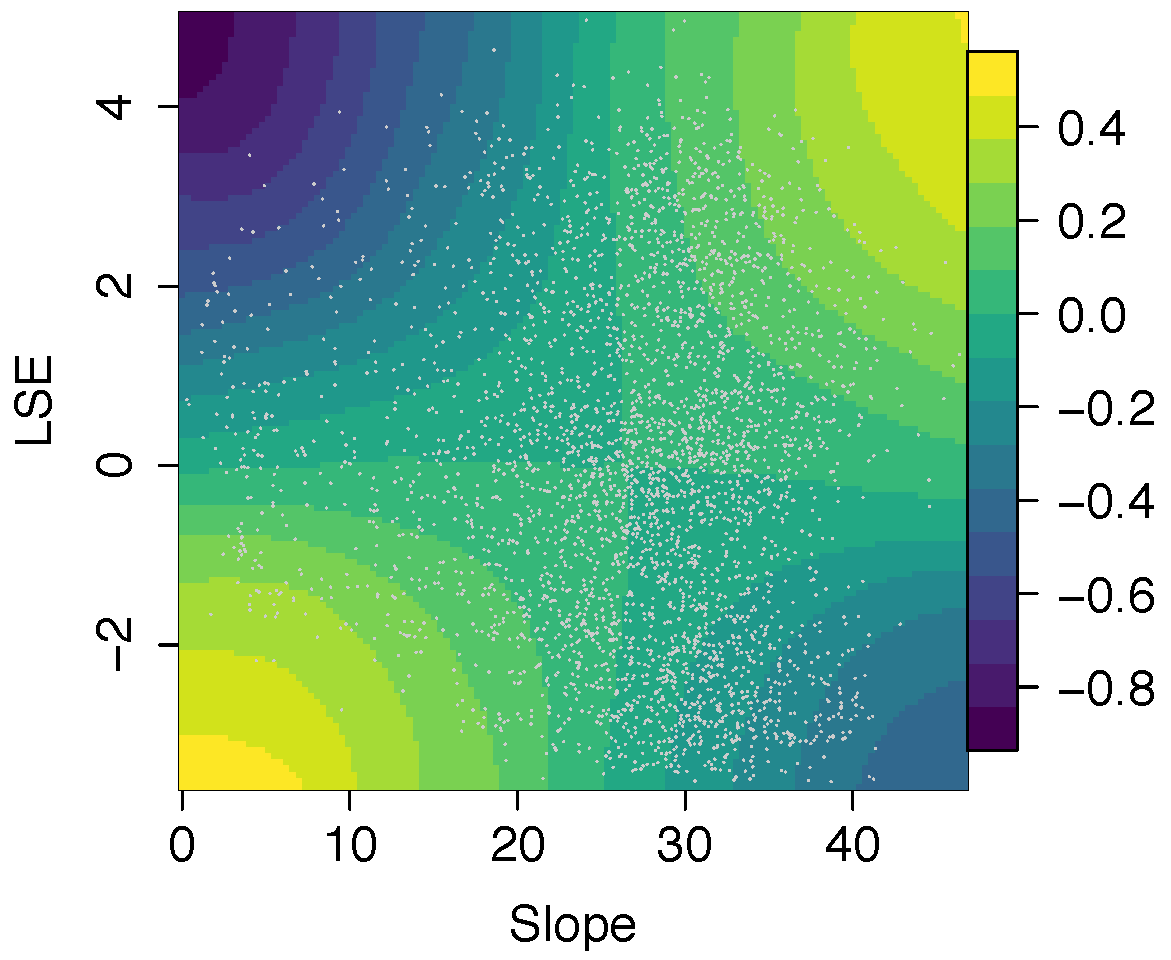} 
	\end{tabular}    
	\caption{Illustration of the interaction structure of Slope Steepness and the precipitation trigger (represented through the LSE) at the SU scale, with Slope Steepness values averaged for each SU. Left column: M4 (SVR). Right column: M5 (P-SVR). First row: smoothed map of SVR coefficient $x_{\mathrm{Slope}}^{\mathrm{CAR}}(\bm s)$. Second row: smoothed map of $\mathrm{Slope}(\bm s)\times x_{\mathrm{Slope}}^{\mathrm{CAR}}(\bm s)$. Small gray dots indicate the LSE values of the $3484$ SUs.}
	\label{fig:slopetrigger}
\end{figure}

\section{Conclusion}
\label{sec:discussion}

For georeferenced landslide events, we have proposed a Bayesian hierarchical point process-based modeling framework to assess landslide hazard at high spatial resolution. Although substantial geomorphological covariate information is available in our case, it remains crucial for spatial prediction and for the interpretation of covariate effects to capture the latent activation pattern induced by the unobserved precipitation trigger. The framework of space-varying regression is appealing and useful, but care is needed to avoid overly complex models, for which model components may be difficult to identify from the data and to interpret intuitively. When a spatial random effect captures the trigger intensity, it is natural to include it as a complement of space-varying regression, such that local covariate effects can be locally removed when the trigger is absent or weak. 
%Our analysis showed that the gain in parsimony and interpretability of the model outweigh the sacrifice of purely linear structures of model parameters. 

The inclusion of  additional latent random effects without spatial dependence in the log-intensity allows us to model small-scale variability,  either at the pixel level (here available at a high $15$m resolution), or at the coarser slope unit (SU) level. This enables capturing overdispersion that would otherwise lead to underestimated local variance of counts arising in the Poisson regression model used to spatially discretize the log-Gaussian Cox process. With our dataset, the inclusion of such i.i.d.\ effects did not substantially improve the baseline model. Rather, information criteria and some of the predictive performance measures indicated that model M1a with pixel-based i.i.d.\ effect was much worse, which may be explained by the large number of additional independent latent variables. This leads to a strong increase in computational complexity of estimation, and may hamper the identification of intermediate-scale spatially structured effects. 

The Bayesian approach is useful to incorporate expert knowledge and well-known physical behavior about the shape of the functional response of processes  to predictor variables into prior models. A large majority of landslides are usually triggered by a specific physical or weather event, such as extreme precipitation in the case of our dataset, and we need spatial random effects too capture the trigger if it has not been observed at high spatial resolution. Moreover, our formulation of the parsimonious space-varying regression is a natural mechanism that allows the model to push the  regression coefficient of Slope Steepness towards $0$ when the trigger is weak. More generally, including such mechanisms into models seems particularly promising for improving interpretability and predictions in cases where a large fraction of the study area experiences only a weak trigger intensity. 

Identifiability problems may arise from the use of several latent components, whose spatial resolution and prior specification allows capturing similar types of variability, and which are not identifiable in a frequentist framework without prior distributions (e.g., spatial random effects, space-varying regression coefficients, and i.i.d.\ effects, all resolved at the SU level). Typically, the Bayesian paradigm explains spatial variation through the component that is most easily ``pushed away'' from its prior towards the posterior shape of the point process intensity. Nevertheless, some confounding between such effects is common. For instance, estimated coefficients of fixed effects for covariates whose values are relatively smooth in space often tend be  slightly smaller in  absolute value in models with spatially resolved random effects. If the ultimate goal is spatial prediction for unobserved areas (e.g., prediction of landslide intensity for unobserved SUs adjacent to the observed ones), we must be careful to avoid a transfer of information from spatially dependent components (such as the latent spatial effect in our model) towards spatially independent effects (such as the i.i.d.\ effect). Our approach  relies on penalized complexity priors where we penalize components rather strongly if they lead to very complex models, which is an appealing solution to cope with sophisticated latent models that include a moderate number of different random effect components. Moreover, we underscore that the  INLA method, combined with modern computing power, provides a convenient toolbox that allows for relatively simple implementation and estimation of very high-dimensional and sophisticated models in the Bayesian framework using latent Gaussian models.

We conclude that model choice is not easy when observed spatial data are discrete, only available through a relatively moderate sample size, not replicated in time, but depend on predictors that may change rapidly even at small spatial scales---in this setting, there usually is not a single ``best" model, and data cannot inform us about the true, complex structure of the intensity function of the point process with high certainty. Therefore, careful construction of a moderate number of candidate models using prior expert knowledge is important.  We recommend that several model diagnostics related to both goodness-of-fit and predictive performance be compared and carefully studied in practice. %\\
%\TC{MENTION MARKED PPS HERE? Raphael: Not sure it's needed...}\\
%\TC{Raphael: Mention that INLA allows to estimate complex models?}\\
%\TC{Raphael: I didn't check carefully the bibliography below, but we should make sure that all papers have been updated, and that we have a consistent style throughout the bibliography (capital letters, journal abbreviations, doi or not, etc.)}\\

%Data of additional landslide characteristics such as the runout length are available for a subarea of the study region containing approximately $1000$ points and for similar datasets over different study regions. We plan to extend the current work to  marked point processes for jointly modeling landslide departure points and  trajectory-related information such as their size. It would be instructive to explore  if such landslide characteristics depend directly on specific components of the intensity model such as the spatial random effect. 

%\bibliographystyle{imsart-nameyear}

\bibliographystyle{apalike}
\bibliography{landslides}

\end{document}